\documentclass[linenumbers]{aastex631}

\usepackage{amsmath}
\usepackage{color}
\usepackage{fancyvrb}
\usepackage{ulem}
\usepackage{graphicx}  % got figures? uncomment this
\usepackage{lineno}

\shorttitle{EstrellaNueva}
\shortauthors{González-Reina et al.}
\begin{document}

\nolinenumbers

\title{EstrellaNueva: an open-source software to study the interactions and detection of neutrinos emitted by supernovae}

\correspondingauthor{O. I. González-Reina}
\email{oiglez@estudiantes.fisica.unam.mx}

\author{O. I. González-Reina}
\affiliation{Instituto de Física, Universidad Nacional Autónoma de México, A.P. 20-364, Ciudad de
México 01000, México
}

\author{J. Rumleskie}
\affiliation{Laurentian University, Department of Physics, 935 Ramsey Lake Road, Sudbury, ON P3E 2C6, Canada}

\author{E. Vázquez-Jáuregui}
\affiliation{Instituto de Física, Universidad Nacional Autónoma de México, A.P. 20-364, Ciudad de
México 01000, México
}
%\end{comment}

%\collaboration{6}{(SNO+ Collaboration)}

%% Note that the \and command from previous versions of AASTeX is now
%% depreciated in this version as it is no longer necessary. AASTeX 
%% automatically takes care of all commas and "and"s between authors names.

%% AASTeX 6.31 has the new \collaboration and \nocollaboration commands to
%% provide the collaboration status of a group of authors. These commands 
%% can be used either before or after the list of corresponding authors. The
%% argument for \collaboration is the collaboration identifier. Authors are
%% encouraged to surround collaboration identifiers with ()s. The 
%% \nocollaboration command takes no argument and exists to indicate that
%% the nearby authors are not part of surrounding collaborations.

%% Mark off the abstract in the ``abstract'' environment. 
\begin{abstract}
\nolinenumbers
Supernovae emit large fluxes of neutrinos which can be detected by detectors on Earth. Future tonne-scale detectors will be sensitive to several neutrino interaction channels, with thousands of events expected if a supernova emerges in the galaxy neighborhood. There is a limited number of tools to study the interaction rates of supernova neutrinos, although a plethora of available supernova models exists. EstrellaNueva is an open-source software to calculate expected rates of supernova neutrinos in detectors using target materials with typical compositions, and additional compositions can be easily added. This software considers the flavor transformation of neutrinos in the supernova through the adiabatic Mikheyev--Smirnov--Wolfenstein effect, and their interaction in detectors through several channels. Most of the interaction cross sections have been analytically implemented, such as neutrino-electron and neutrino-proton elastic scattering, inverse beta decay, and coherent elastic neutrino-nucleus scattering. This software provides a link between supernova simulations and the expected events in detectors by calculating fluences and event rates to ease any comparison between theory and observation. It provides a simple and standalone tool to explore many physics scenarios offering an option to add analytical cross sections and define any target material.
\end{abstract}

%% Keywords should appear after the \end{abstract} command. 
%% The AAS Journals now uses Unified Astronomy Thesaurus concepts:
%% https://astrothesaurus.org
%% You will be asked to selected these concepts during the submission process
%% but this old "keyword" functionality is maintained in case authors want
%% to include these concepts in their preprints.
%\keywords{Classical Novae (251) --- Ultraviolet astronomy(1736) --- History of astronomy(1868) --- Interdisciplinary astronomy(804)}

%% From the front matter, we move on to the body of the paper.
%% Sections are demarcated by \section and \subsection, respectively.
%% Observe the use of the LaTeX \label
%% command after the \subsection to give a symbolic KEY to the
%% subsection for cross-referencing in a \ref command.
%% You can use LaTeX's \ref and \label commands to keep track of
%% cross-references to sections, equations, tables, and figures.
%% That way, if you change the order of any elements, LaTeX will
%% automatically renumber them.
%%
%% We recommend that authors also use the natbib \citep
%% and \citet commands to identify citations.  The citations are
%% tied to the reference list via symbolic KEYs. The KEY corresponds
%% to the KEY in the \bibitem in the reference list below. 

\section{Introduction}

Supernovae (SNe) are one of the most intriguing and least known phenomena in the universe. In this process, approximately $3\times 10^{53}$ erg of energy are emitted in form of neutrinos of all species in the energy range of MeV, which represent $99\%$ of the total released energy  \citep{Lunardini:2005jf}. Questions that are still open in neutrino physics such as their Dirac or Majorana nature \citep{Barranco_2020}, their masses \citep{Lu_2015}, the hierarchy between the masses \citep{PhysRevD.85.085031, PhysRevD.62.033007}, and the neutrino magnetic moment \citep{PhysRevD.67.023004}, are the motivation of current scientific research and could be solved by studying the high neutrino fluxes arriving to Earth from nearby SNe events. Current and future neutrino detectors, such as SNO+ \citep{Albanese_2021}, Super-Kamiokande \citep{Ikeda_2007}, KamLAND \citep{PhysRevLett.90.021802}, HALO \citep{Duba_2008}, JUNO \citep{An_2016}, and Hyper-Kamiokande \citep{Hyper-Kamiokande:2018ofw}, are capable of detecting the neutrino flux coming from a supernova (SN) explosion in the Milky Way galaxy or the galactic neighborhood. Such neutrinos can also be detected on dark matter detectors through the process coherent elastic neutrino-nucleus scattering (CE$\nu$NS) \citep{PhysRevD.68.023005, PhysRevD.94.103009}. This represents an invaluable opportunity for exploring neutrino properties and SN physics. On the other hand, there is a community of researchers dedicated to simulating the SN neutrino flux~\citep{web:Garching}, considering many parameters and variables. The EstrellaNueva software has been created to provide a link between the SN simulations and the expected observed signal in detectors. EstrellaNueva is a flexible and simple standalone tool that is valuable to promptly perform several studies when a SN takes place in the galaxy neighborhood. 

The software package has several SN models implemented, the adiabatic Mikheyev--Smirnov--Wolfenstein (MSW) \citep{PhysRevD.62.033007, Giunti:1053706, Xu_2014} effect for neutrino flavor transformation in the SN matter, and the analytical cross sections of the principal detection channels in current and future detectors. Another important and unique feature of this software is the option to design any target material. EstrellaNueva allows calculating the fluences of the neutrinos coming from SNe, interaction rates, and total number of interactions with the target particles of the detectors. These characteristics are valuable for performing comparisons of events observed by detectors that are either currently operating or under design and construction.

In addition to EstrellaNueva, other software tools to study supernova neutrinos exist, but are generally not as simple to use and require additional software packages. For example, SNEWPY \citep{SNEWS:2021ezc} requires an event rate calculator or an event generator such as sntools \citep{Migenda2021}, while SNOwGLoBES \citep{snowglobes} requires the GLoBES \citep{HUBER2005195,HUBER2007432} software to perform the event rate calculations. EstrellaNueva provides all in a single package without requiring any other interface, and also provides an analytical implementation of the cross section functions and the option to input target materials with any chemical composition. This advantage makes possible the implementation of any SN model and any neutrino interaction in the detector. The modular implementation also makes it suitable to study other neutrino flavor scenarios, not only in the SN, but also when the neutrino propagates on Earth. The software currently considers flavor transformation via the adiabatic MSW effect in the SN.
 
In this manuscript, the EstrellaNueva software is presented with a description of the SN models used and the neutrino physics implemented, with some examples of its flexibility and simplicity. The followings sections are organized as follows: Section~\ref{sec:code_structure} shows the software structure; Section~\ref{sec:Flavor transformation and expected interaction rate} describes a general overview of the implemented physics; Section~\ref{sec:supernova_models} shows the implemented SN models; Section~\ref{sec:detection_channels} presents the analytically implemented cross sections of the detection channels and the implemented nuclear form factors for the case of CE$\nu$NS interaction; Section~\ref{sec:usage_examples} shows two examples of the configuration file setup; finally Section~\ref{sec:conclusions} presents the conclusions.\\ 

\section{Software structure}
\label{sec:code_structure}

EstrellaNueva is open-source software written in Python designed to complement the supernova simulations, by calculating the interaction rates as well as the number of events in detectors. It was built using NumPy \citep{Numpy} and SciPy \citep{Scipy}: two well-optimized Python modules that offer excellent tools to process large volumes of data and perform scientific calculations. It also uses the Python module Mathplotlib \citep{Hunter:2007} to plot the output functions as well as some of the implemented functions such as cross sections and nuclear form factors. The software is available in the Zenodo repository:  \url{https://doi.org/10.5281/zenodo.6354850} \citep{gonzalez_reina_o_i_2022_6354850}.

The software package, stored in the directory ``EstrellaNueva", contains the following items: 
\begin{itemize}
    \item The main file ``EstrellaNueva.py": main script to execute the software. This Python script uses Python 3 as interpreter, from the directory ``EstrellaNueva". The command line for execution is: ``\texttt{python3 EstrellaNueva.py}". 
    \item The configuration file ``config.json": JSON file for setting the configuration parameters, before executing the main script.   
    \item The sub-directory ``data": contains the data of the supernova models, as well as some of the implemented cross sections which were taken from the SNOwGLoBES \citep{snowglobes} repository in form of data files. These cross sections correspond to the interactions with ${}^{12}$C, ${}^{16}$O, and ${}^{208}$Pb.
    \item The sub-directory ``out": will contain the output files once the code is executed. The output consists of data files with the number of interactions per channel, interaction rates, and fluences. 
    \item The sub-directory ``src": contains the EstrellaNueva libraries. The cross sections implemented analytically are located in this directory.
    \item The sub-directory ``doc": contains the user manual.
\end{itemize}

The input requested by the code is:
\begin{itemize}
    \item the supernova model,
    \item the neutrino mass ordering \footnote{The neutrino mass ordering is considered for the adiabatic MSW effect in the supernova matter. Flavor transformations can also be turned off.}  
    \item the distance to the supernova, 
    \item the mass of the active volume of the detector and chemical composition of the material, and
    \item the detection channels; 
\end{itemize}
and the output is: 
\begin{itemize}
    \item the interaction rates with respect to the incoming neutrino energy and time, in addition to the recoil particle energy, for the case of elastic scattering interactions, 
    \item the fluences, and 
    \item the total number of interactions.
\end{itemize}
    
\section{Flavor transformation and expected interaction rate}\label{sec:Flavor transformation and expected interaction rate}

Most SN simulations report the time dependence of the neutrino flux at the neutrino-sphere. At this location inside the SN, it is assumed neutrinos can travel freely, without further diffusing with the SN matter due to their small interaction cross sections. The neutrino flux is assumed to be quasi-thermal in this sphere and can be parameterized by the following expression as a function of the neutrino energy $E$ and the SN time $t$ (the elapsed time taking the origin at the core bounce) \citep{Keil_2003,PhysRevD.85.085031}:
\begin{equation}
F_{\nu}^0(E,t) =\mathcal{L}_{\nu}(t)
\frac{\left(1+\beta_{\nu}(t)\right)^{1+\beta_{\nu}(t)}}{\Gamma\left(1+\beta_{\nu}(t)\right)}\frac{E^{\beta_{\nu}(t)}}{\left(\left<E_{\nu}\right>(t)\right)^{\beta_{\nu}(t)+2}}\times\text{exp}\left[-\left(\beta_{\nu}(t)+1\right)\frac{E}{\left<E_{\nu}\right>(t)}\right],
\label{eq:flujo1} 
\end{equation}
where $\mathcal{L}_{\nu}$ is the luminosity, $\left<E_{\nu}\right>$ is the neutrino mean energy, $\beta_{\nu}$ is the so-called energy-shape parameter, and $\nu$ represents the neutrino states $\nu_e$, $\nu_{\mu}$, $\nu_{\tau}$, $\bar{\nu}_e$, $\bar{\nu}_{\mu}$, and $\bar{\nu}_{\tau}$. The energy-shape parameter is given by 
\begin{equation}
\beta_{\nu}(t)=\frac{2\left(\left<E_{\nu}\right>(t)\right)^2-\left<E_{\nu}^2\right>(t)}{\left<E^2_{\nu}\right>(t)-\left(\left<E_{\nu}\right>(t)\right)^2},
\label{eq:beta_en_t}
\end{equation}   
which quantifies how close the neutrino spectrum is to the black-body spectrum, as a function of time \citep{PhysRevD.85.085031}. The primary fluxes at the neutrino-sphere $F^0_{\nu}$ will be affected by neutrino flavor transformation on their path through the SN matter. One of the most common and accepted scenarios is the adiabatic MSW effect, which has been implemented in the EstrellaNueva framework.\\

\subsection{Adiabatic MSW effect}

An ultrarelativistic left-handed neutrino, born at the neutrino-sphere, with flavor $\alpha$ ($\alpha = e, \mu, \tau$) and momentum $\vec{p}$, is described by the flavor eigenstate 
\begin{equation}
    |\nu_{\alpha}\rangle = \sum_{k}^{}U^*_{\alpha k}|\nu_k \rangle, 
\end{equation}
where $U$ is the neutrino mixing matrix:
\begin{equation}
    U = \left(\begin{array}{ccc}
         c_{12}c_{13}&s_{12}c_{13}&s_{13}e^{-i\delta} \\
         -s_{12}c_{23}-c_{12}s_{23}s_{13}e^{i\delta}&c_{12}c_{23}-s_{12}s_{23}s_{13}e^{i\delta}&s_{23}c_{13}\\
         s_{12}s_{23}-c_{12}s_{23}s_{13}e^{i\delta}&-c_{12}s_{23}-s_{12}c_{23}s{13}e^{i\delta}&c_{23}c_{13}
    \end{array}\right),
\end{equation}
with $s_{ij}=\sin \theta_{ij}$, $c_{ij}=\cos \theta_{ij}$. The $\theta_{ij}$ are mixing angles and $\delta$ is a Dirac CP\footnote{CP is the combined charge conjugation (C) and parity (P) symmetry.} violating phase~\citep{deSalas:2020pgw}. The symbol $|\nu_k \rangle$, where $k=1,2,3$, represents an eigenstate of the vacuum Hamiltonian $\mathcal{H}_0$:
\begin{equation}
    \mathcal{H}_0|\nu_k \rangle = E_k|\nu_k\rangle, \hspace{1cm}\text{with}\hspace{1cm}E_k = \sqrt{\vec{p}^2 + m_k^2}.
\end{equation}
The total Hamiltonian in matter is 
\begin{equation}
    \mathcal{H} = \mathcal{H}_0 + \mathcal{H}_I,\hspace{1cm}\text{with}\hspace{1cm} \mathcal{H}_I|\nu_{\alpha}\rangle = V_{\alpha}|\nu_{\alpha}\rangle,
\end{equation}
where $V_{\alpha}$ is the effective potential experienced by the ultrarelativistic left-handed flavor neutrino: 
\begin{equation}
  V_{\alpha} = V_{CC}\delta_{\alpha e} + V_{NC} =  \sqrt{2}G_F\left(n_e\delta_{\alpha e} - \frac{1}{2}n_n\right).
  \label{eq:effective_potential}
\end{equation}
In Equation (\ref{eq:effective_potential}), $\delta_{\alpha e}$ is the Kronecker delta, $V_{CC}$ and $V_{NC}$ are the effective potentials due to the charge current (CC) and neutral current (NC) interactions, and $n_e$ and $n_n$ are the electron and neutron density of the medium, respectively. In the Schrödinger picture, a neutrino state with initial flavor $\alpha$ obeys the evolution equation 
\begin{equation}
	i\frac{\text{d}}{\text{d}t}|\nu_{\alpha}(t)\rangle = \mathcal{H}|\nu_{\alpha}(t)\rangle, \hspace{1cm}\text{with} \hspace{1cm} |\nu_{\alpha}(0)\rangle = |\nu_{\alpha}\rangle.
\end{equation}
For ultrarelativistic neutrinos the following approximation is valid:
\begin{equation}
	    E_k \simeq E + \frac{m_k^2}{2E}, \hspace{1cm} p \simeq E, \hspace{1cm} t\simeq x,
\end{equation}
where $x$ is the distance from the neutrino-sphere. Neglecting a common phase term, the evolution equation for the amplitude of the transition $\nu_{\alpha} \overrightarrow{} \nu_{\beta}$, $\psi_{\alpha \beta} = \langle \nu_{\beta}|\nu_{\alpha}(t)\rangle$ (with $\beta = e, \mu, \tau$), can be written in matrix form as \citep{Giunti:1053706}
\begin{equation}
	i\frac{\text{d}}{\text{d}x}\Psi_{\alpha} = \mathcal{H}_F \Psi_{\alpha}, 
\end{equation}
where $\mathcal{H}_F$ is the effective Hamiltonian in the flavor basis:
\begin{equation}
	\mathcal{H}_F = \frac{1}{2E}\left(U \mathbb{M}^2 U^{\dagger} + \mathbb{A} \right).
\end{equation}
In the case of three-neutrino mixing,
\begin{equation}
\Psi_{\alpha} = \left(\begin{array}{c}
	    \psi_{\alpha e}\\
	    \psi_{\alpha \mu}\\
	    \psi_{\alpha \tau}
	  \end{array}\right), \hspace{.6cm} 
	    \mathbb{M}^2 = \left(\begin{array}{ccc}
	    0 & 0 & 0\\
	    0 & \Delta m_{21}^2 & 0\\
	    0 & 0 & \Delta m_{31}^2
	    \end{array}\right), \hspace{.6cm}
	    \mathbb{A} = \left(\begin{array}{ccc}
	    A_{CC} & 0 & 0\\
	    0 & 0 & 0\\
	    0 & 0 & 0
	    \end{array}\right),
	    \end{equation}
where $\Delta m^2_{21}$ and $\Delta m^2_{31}$ are the squared neutrino mass splittings, and   
	    \begin{equation}
	    A_{CC}(x) \equiv 2EV_{CC} (x) = 2\sqrt{2}E G_F n_e(x).
	    \end{equation}
An analogous mathematical development can be obtained for anti-neutrinos, where $A_{CC}(x) = -2\sqrt{2}E G_F n_e(x)$ due to the change in the sign of $V_{CC}$. More details about this description can be found in \citep{Giunti:1053706}. 

The primary fluxes coming from the neutrino-sphere are $F_{\nu_e}^0$, $F_{\bar{\nu}_e}^0$, and $F_{\nu_x}^0$, where $\nu_x=\nu_{\mu}, \nu_{\tau}, \bar{\nu}_{\mu}, \bar{\nu}_{\tau}$. Because non-electron type neutrinos have the same interactions inside the SN, their primary fluxes are expected to be approximately equal \citep{PhysRevD.62.033007}. It is useful to use the equality $F_{\nu_{\mu}}^0=F_{\nu_{\tau}}^0 = F_{\nu_x}^0$ and perform a rotation in the ($|\nu_{\mu}\rangle, |\nu_{\tau}\rangle$) plane changing the basis, ($|\nu_e\rangle, |\nu_{\mu}\rangle, |\nu_{\tau}\rangle) \rightarrow (|\nu_e\rangle, |\nu_{\mu}^{'}\rangle,|\nu_{\tau}^{'}\rangle)$:
\begin{equation}
	|\nu^{'}_{\mu}\rangle = \cos \theta_{23} |\nu_{\mu}\rangle - \sin \theta_{23} |\nu_{\tau}\rangle, \hspace{1cm}
	|\nu^{'}_{\tau}\rangle = \sin \theta_{23} |\nu_{\mu}\rangle + \cos \theta_{23} |\nu_{\tau}\rangle,
\end{equation}
as described in \citep{V_n_nen_2011}. This rotation ensures that $F_{\nu_x}^0=F_{\nu_{\mu}^{'}}^0 = F_{\nu_{\tau}^{'}}^0$ and diagonalizes the ($\nu_{\mu}, \nu_{\tau}$) submatrix of $\mathcal{H}_F$: 
\begin{equation}
	\mathcal{H}_F^{rot} = \frac{1}{2E}\left( \begin{array}{ccc}
	m_{ee}^2 + A_{CC} & m_{e \mu^{'}}^2 & m_{e \tau^{'}}^2\\
	m_{e \mu^{'}}^2 & m_{\mu^{'}\mu^{'}}^2 & 0\\
	m_{e \tau^{'}}^2 & 0 & m_{\tau^{'}\tau^{'}}^2
	\end{array}\right).
\end{equation}

\begin{figure}
    \centering
    \includegraphics[width=.9\textwidth]{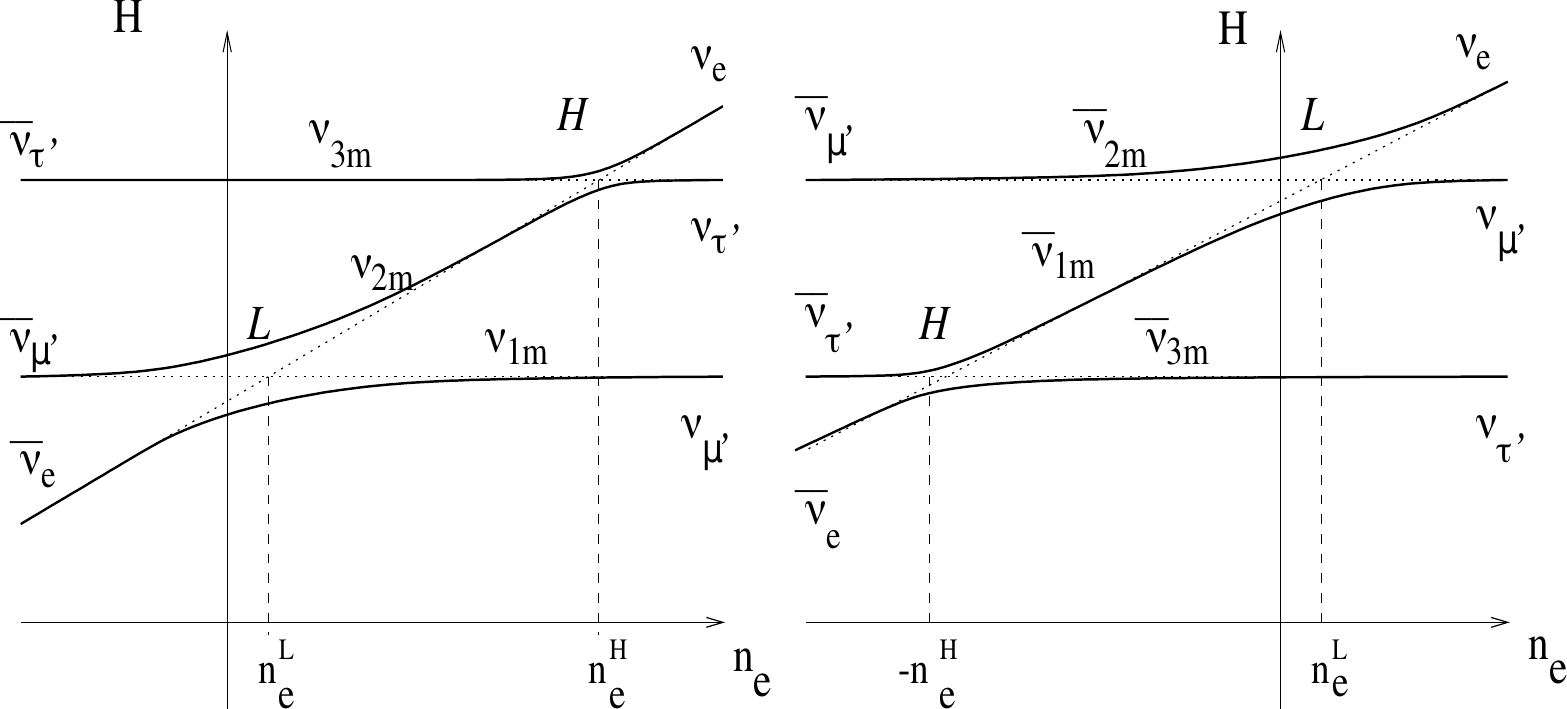}
    \caption{Level crossing diagrams for normal (left) and inverted (right) mass ordering of the neutrinos. Solid lines show the eigenvalues of the effective Hamiltonian $\mathcal{H}_F^{rot}$ in the effective matter basis as function of the electron density $n_e$. Dashed lines correspond to energies of the flavor levels. The region with $n_e>0$ refers to neutrinos, while the one with $n_e<0$ to anti-neutrinos. \textit{L} denotes the resonance region at low $n_e$ and \textit{H} the resonance region at high $n_e$. Figure reprinted with permission from \citep{PhysRevD.62.033007} (\url{https://doi.org/10.1103/PhysRevD.62.033007}).}
    \label{fig:level_crossing_diagrams}
\end{figure}

At high matter densities, close to the neutrino-sphere, the other off-diagonal elements of $\mathcal{H}_F^{rot}$ can be neglected. This means that the flavor eigenstates $(|\nu_e \rangle,|\nu^{'}_{\mu} \rangle,|\nu{'}_{\tau}\rangle )$ coincide with the effective matter eigenstates $(|\nu_{1m} \rangle,|\nu_{2m} \rangle, |\nu_{3m}\rangle)$ when they are produced. The rotated effective Hamiltonian $\mathcal{H}_{F}^{rot}$ allows one to construct the level diagrams shown in Figure \ref{fig:level_crossing_diagrams}, for a given neutrino energy value \citep{PhysRevD.62.033007}. During the first second of the burst of neutrinos from a SN, in typical SN density profiles (model independent), the neutrino evolution can be considered adiabatically \citep{Xu_2014}. This means that the transition probability between effective matter eigenstates can be neglected. Then, a neutrino born in an effective matter eigenstate will remain in this state during its path through the SN matter, until reaching the vacuum, where $|\nu_{km}\rangle \rightarrow |\nu_k \rangle$. Therefore, taking into account the mixing of the neutrinos and following the level diagrams shown in Figure \ref{fig:level_crossing_diagrams}, the neutrino fluxes in vacuum, considering the neutrino normal mass ordering (NO), can be written as a linear combination of the primary fluxes at the neutrino-sphere, 
\begin{equation}
	\begin{split}
	 &F_{\nu_e}=\left|U_{e1}\right|^2 F_{\nu_x}^0 + \left|U_{e2}\right|^2 F_{\nu_x}^0 + \left|U_{e3}\right|^2 F_{\nu_e}^0,\\
	 &F_{\nu_{\mu}}=\left|U_{\mu 1}\right|^2 F_{\nu_x}^0 + \left|U_{\mu 2}\right|^2 F_{\nu_x}^0 + \left|U_{\mu 3}\right|^2 F_{\nu_e}^0,\\
	 &F_{\nu_{\tau}}=\left|U_{\tau 1}\right|^2 F_{\nu_x}^0 + \left|U_{\tau 2}\right|^2 F_{\nu_x}^0 + \left|U_{\tau 3}\right|^2 F_{\nu_e}^0,\\
	 &F_{\bar{\nu}_e}=\left|U_{e1}\right|^2 F_{\bar{\nu}_e}^0 + \left|U_{e2}\right|^2 F_{\nu_x}^0 + \left|U_{e3}\right|^2 F_{\nu_x}^0,\\
	 &F_{\bar{\nu}_{\mu}}=\left|U_{\mu 1}\right|^2 F_{\bar{\nu}_e}^0 + \left|U_{\mu 2}\right|^2 F_{\nu_x}^0 + \left|U_{\mu 3}\right|^2 F_{\nu_x}^0,\\
	 &F_{\bar{\nu}_{\tau}}=\left|U_{\tau 1}\right|^2 F_{\bar{\nu}_e}^0 + \left|U_{\tau 2}\right|^2 F_{\nu_x}^0 + \left|U_{\tau 3}\right|^2 F_{\nu_x}^0.
	 \end{split}
\label{eq:oscillated_fluxes_NO}
\end{equation}
Similarly, for the neutrino inverted mass ordering (IO) the neutrino fluxes in vacuum are 
\begin{equation}
	\begin{split}
	 &F_{\nu_e}=\left|U_{e1}\right|^2 F_{\nu_x}^0 + \left|U_{e2}\right|^2 F_{\nu_e}^0 + \left|U_{e3}\right|^2 F_{\nu_x}^0,\\
	 &F_{\nu_{\mu}}=\left|U_{\mu 1}\right|^2 F_{\nu_x}^0 + \left|U_{\mu 2}\right|^2 F_{\nu_e}^0 + \left|U_{\mu 3}\right|^2 F_{\nu_x}^0,\\
	 &F_{\nu_{\tau}}=\left|U_{\tau 1}\right|^2 F_{\nu_x}^0 + \left|U_{\tau 2}\right|^2 F_{\nu_e}^0 + \left|U_{\tau 3}\right|^2 F_{\nu_x}^0,\\
	 &F_{\bar{\nu}_e}=\left|U_{e 1}\right|^2 F_{\nu_x}^0 + \left|U_{e 2}\right|^2 F_{\nu_x}^0 + \left|U_{e 3}\right|^2 F_{\bar{\nu}_e}^0,\\
	 &F_{\bar{\nu}_{\mu}}=\left|U_{\mu 1}\right|^2 F_{\nu_x}^0 + \left|U_{\mu 2}\right|^2 F_{\nu_x}^0 + \left|U_{\mu 3}\right|^2 F_{\bar{\nu}_e}^0,\\
	 &F_{\bar{\nu}_{\tau}}=\left|U_{\tau 1}\right|^2 F_{\nu_x}^0 + \left|U_{\tau 2}\right|^2 F_{\nu_x}^0 + \left|U_{\tau 3}\right|^2 F_{\bar{\nu}_e}^0.
	 \end{split}
\label{eq:oscillated_fluxes_IO}
\end{equation}

\subsection{Interaction rates}

The neutrino interaction rates in a detector on Earth are determined for the flavour fluxes $F_{\nu}$ emitted by the SN as given in Equations (\ref{eq:oscillated_fluxes_NO}) and (\ref{eq:oscillated_fluxes_IO}). The number of interactions per unit time and energy is 
\begin{equation}
\frac{d^2N_{\nu}}{dEdt}(E,t)= \frac{N_t}{4\pi d^2}\sigma_{\nu}(E)F_{\nu}(E,t),
\label{eq:event_rate_E_t}
\end{equation}
where $d$ is the distance to the SN, $N_t$ the number of target particles in the active volume of the detector, and $\sigma_{\nu}$ is the total cross section as a function of the incoming neutrino energy for a given interaction channel. Here is useful to define a magnitude so-called fluence, which is the time-integrated flux at Earth and provides the energy spectrum of the incoming neutrinos: 
\begin{equation}
	\lambda_{\nu}(E) = \frac{1}{4\pi d^2}\int_{t_0}^{t_{end}}F_{\nu}(E,t)dt.
	\label{eq:fluences}
\end{equation}
In Equation (\ref{eq:fluences}), $(t_0; t_{end})$ is the time interval of the SN process.
Integrating Equation (\ref{eq:event_rate_E_t}), the interaction rates with respect to $E$ and $t$ are
\begin{equation}
\frac{dN_{\nu}}{dE}(E)=\int_{t_0}^{t_{end}}dt\left[\frac{d^2N_{\nu}}{dEdt}(E,t)\right] = N_t\sigma_{\nu}(E)\lambda_{\nu}(E),
\label{eq:event_rate_E}
\end{equation}
and
\begin{equation}
\frac{dN_{\nu}}{dt}(t) = \int_{0}^{\infty}dE\left[\frac{d^2N_{\nu}}{dEdt}(E,t)\right], 
\end{equation} 
respectively.

In the case of elastic scattering interactions, like neutrino-electron ($\nu-e$) elastic scattering, neutrino-proton ($\nu-p$) elastic scattering, and CE$\nu$NS, it is possible to calculate the event rates with respect to the recoil particle energy $T$. The number of interactions per unit time, unit energy, and unit recoil energy is given by 
\begin{equation}
\frac{d^3N_{\nu}}{dEdTdt}(E, T, t)= \frac{N_t}{4\pi d^2}\frac{d\sigma_{\nu}}{dT}(E, T)F_{\nu}(E,t),
\label{eq:event_rate_E_T_t}
\end{equation}    
where $\frac{d\sigma_{\nu}}{dT}$ is the differential cross section with respect to $T$, which is related to the total cross section as follows:
\begin{equation}
\sigma_{\nu}(E) = \int_0^{T_{max}(E)}\frac{d\sigma_{\nu}}{dT}(E,T)dT,
\label{eq:maximun_recoil_energy}
\end{equation} 
where 
\begin{equation}
T_{max}(E)= \frac{2E^2}{2E+m},
\end{equation}
is the maximum energy of the recoil particle, 
and $m$ its mass. From Equation (\ref{eq:event_rate_E_T_t}), the interaction rate with respect to $T$ is obtained as follows,
\begin{equation}
\frac{dN_{\nu}}{dT}(T) = \int_{t_0}^{t_{end}}dt\int_{E_{min}(T)}^{\infty}dE\left[\frac{d^3N_{\nu}}{dEdTdt}(E, T, t)\right] = N_t\int_{E_{min}(T)}^{\infty}dE\left[\frac{d\sigma_{\nu}}{dT}(E,T)\lambda_{\nu}(E)\right], 
\end{equation}
where  
\begin{equation}
E_{min}(T) = \frac{T + \sqrt{T(T+2m)}}{2}, 
\end{equation}
is the minimum energy required by the incoming neutrino.
Finally, the total number of interactions is 
\begin{equation}
N_{\nu} = \int_{0}^{\infty} dE \left [\frac{dN_{\nu}}{dE}(E)\right] = \int_{t_0}^{t_{end}}dt\left[\frac{dN_{\nu}}{dt}(t)\right]\\
= \int_{0}^{\infty} dT \left[\frac{dN_{\nu}}{dT}(T)\right].
\end{equation}

\section{Supernova Models}
\label{sec:supernova_models}

\begin{figure}[htbp]
	\gridline{\fig{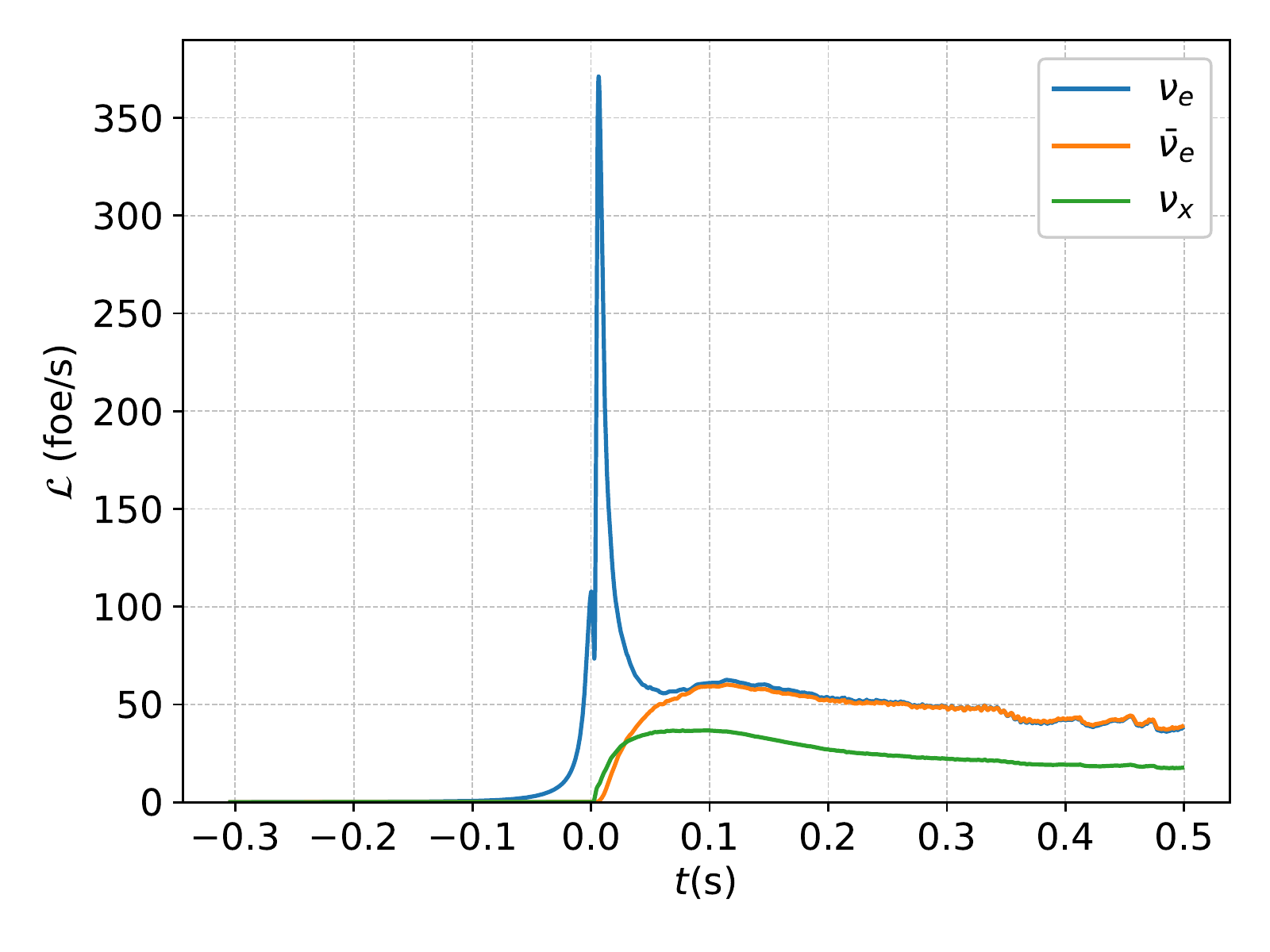}{0.5\textwidth}{(a)}
    \fig{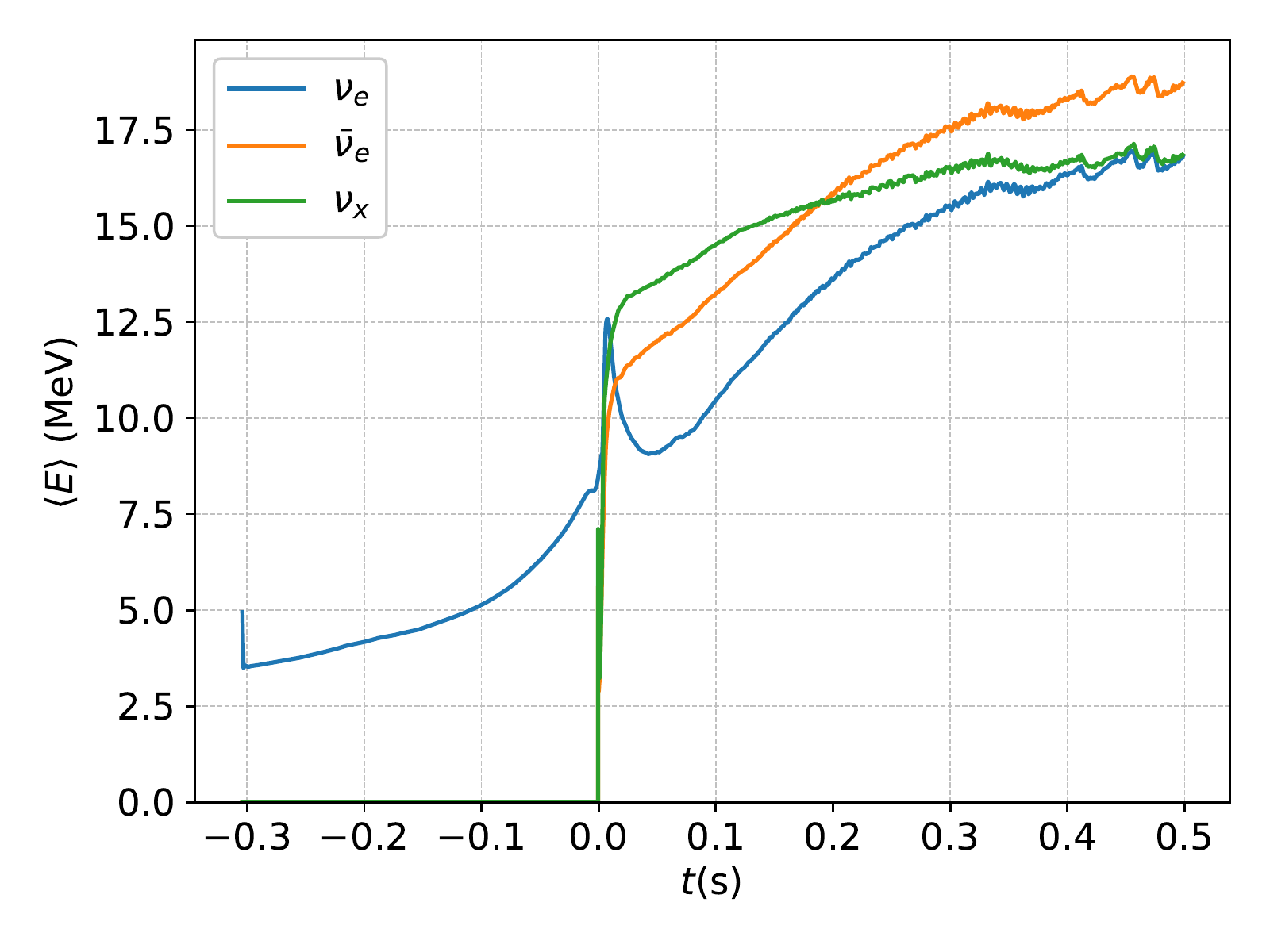}{0.5\textwidth}{(b)}
    }
    \gridline{\fig{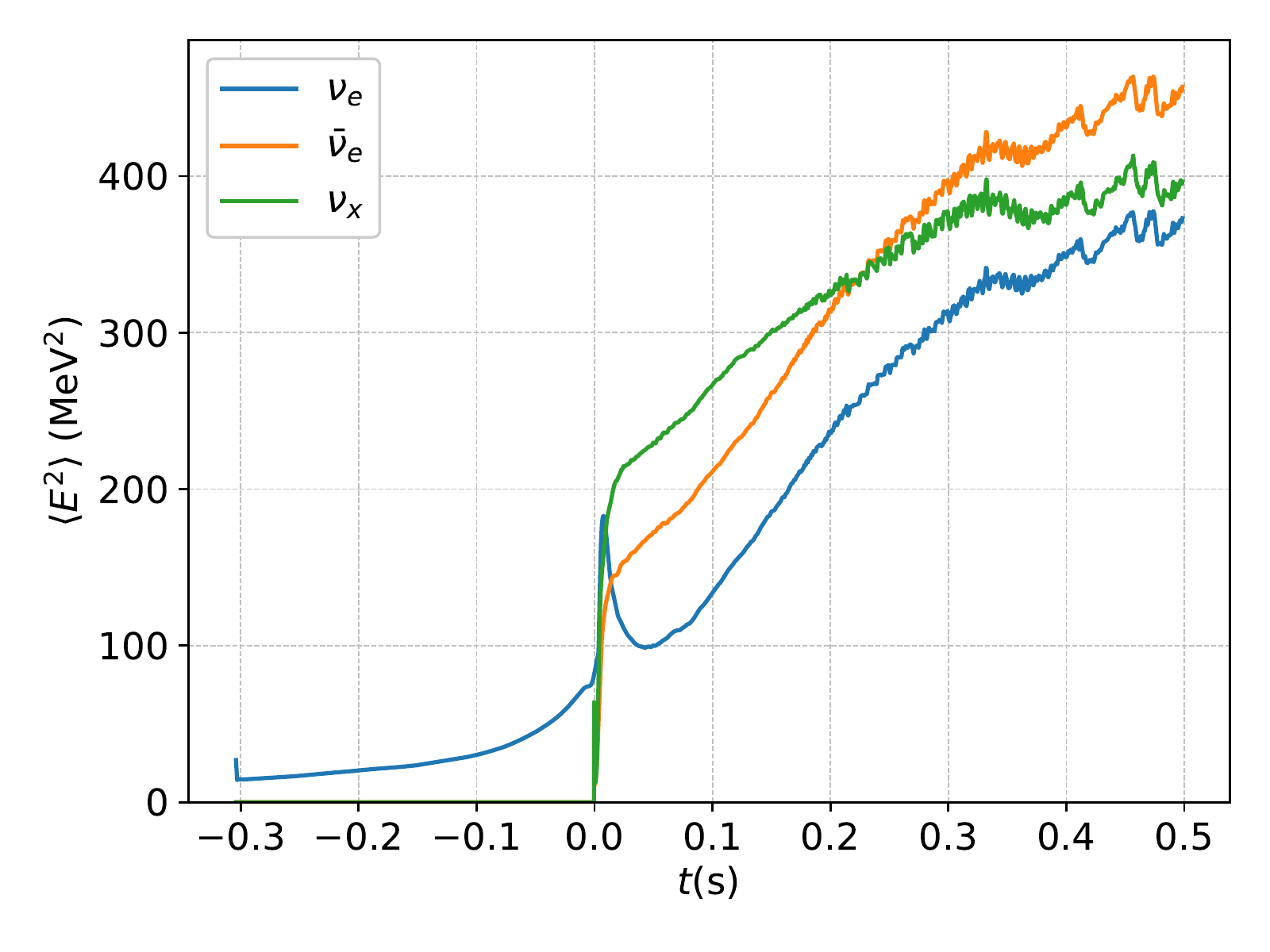}{0.5\textwidth}{(c)}
    \fig{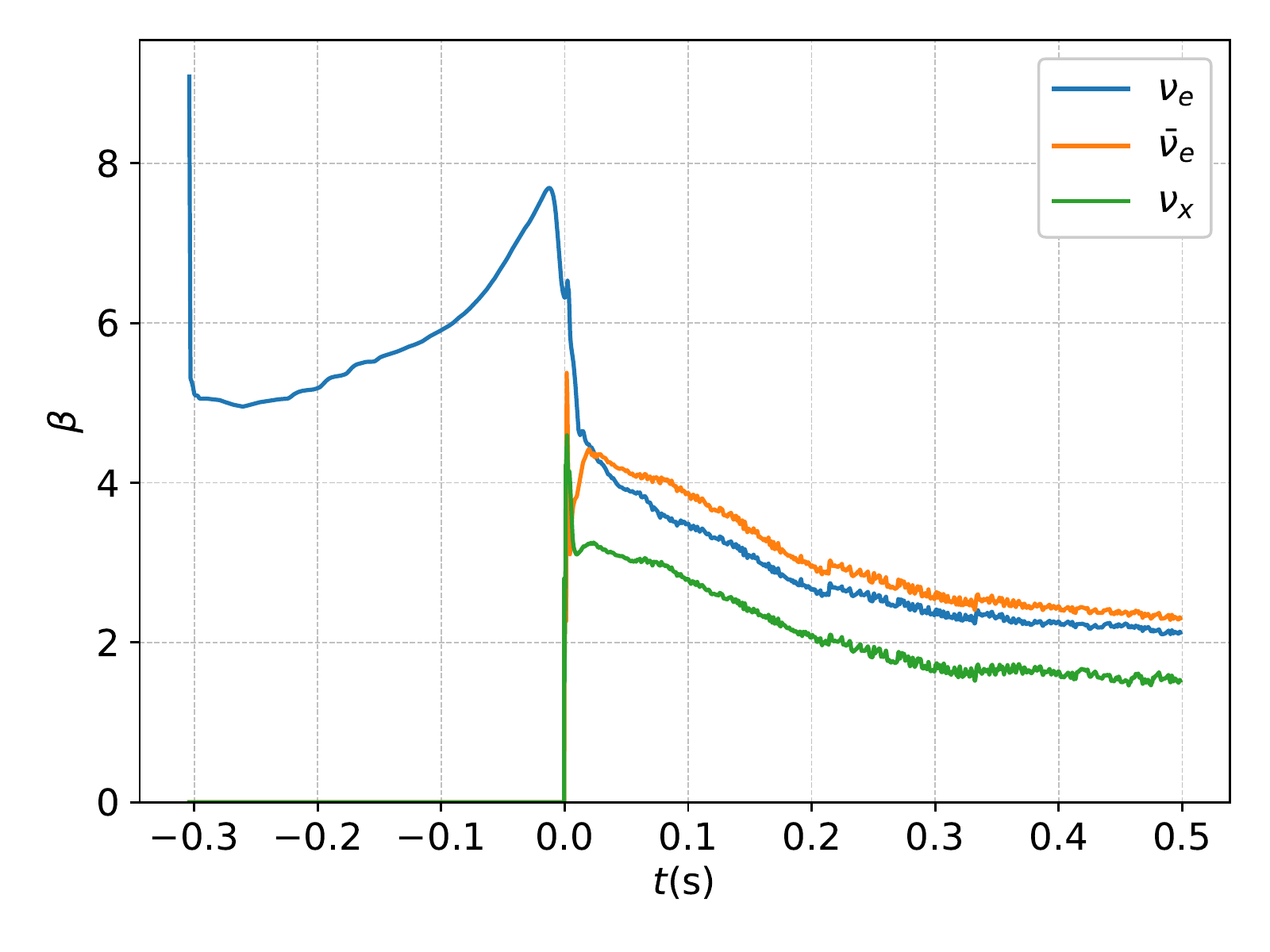}{0.5\textwidth}{(d)}
    }
	\caption{Time dependence of the primary flux parameters corresponding to the SN model LS220-s15.0 \citep{web:Garching}. (a) Luminosity, (b) mean energy, (c) mean squared energy, and (d) energy-shape parameter, which was calculated following Equation (\ref{eq:beta_en_t}).}
	\label{fig:SN_model}  
\end{figure}

\begin{figure}[htbp]
	\gridline{\fig{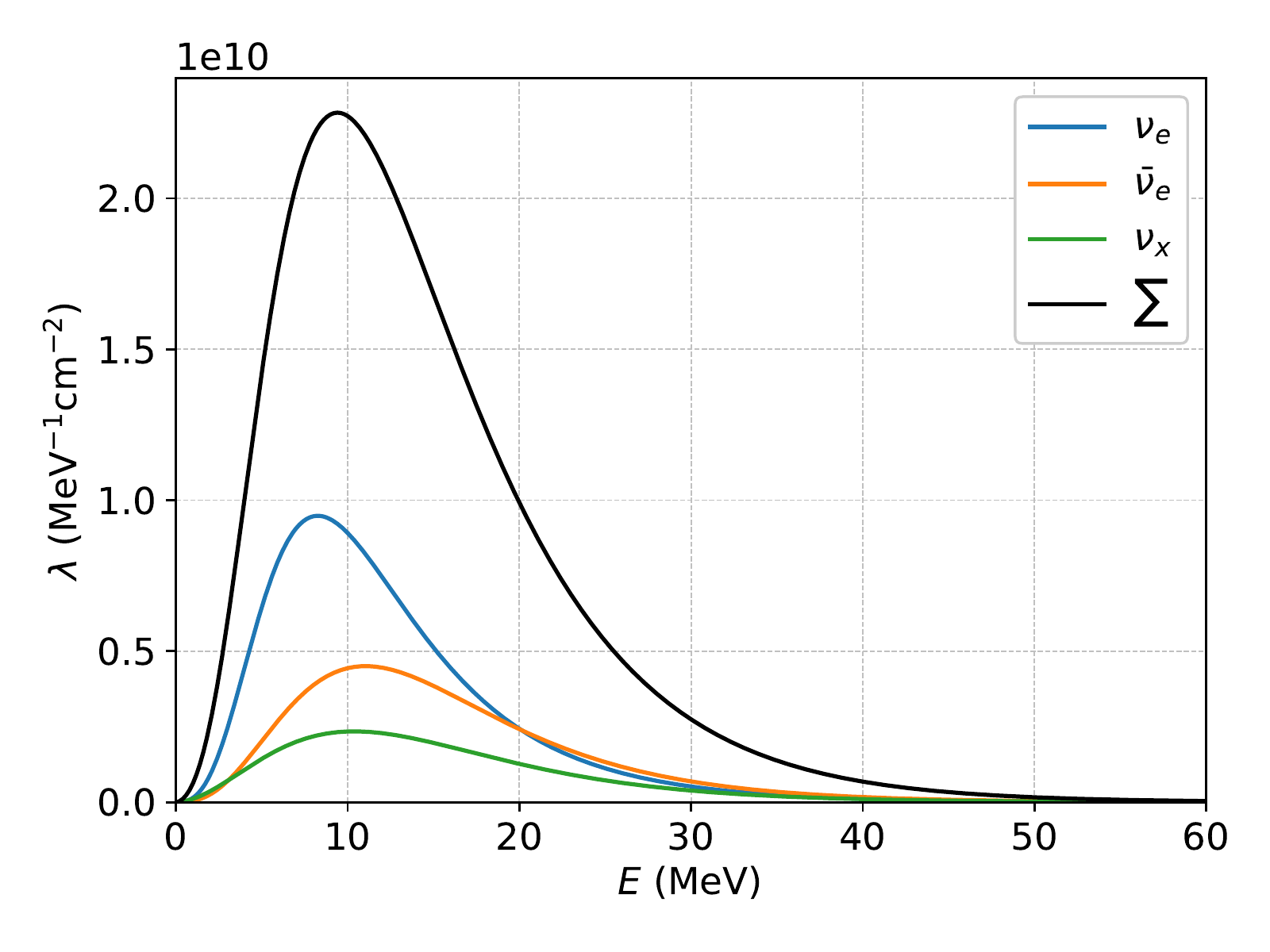}{0.5\textwidth}{(a)}
    }
    \gridline{\fig{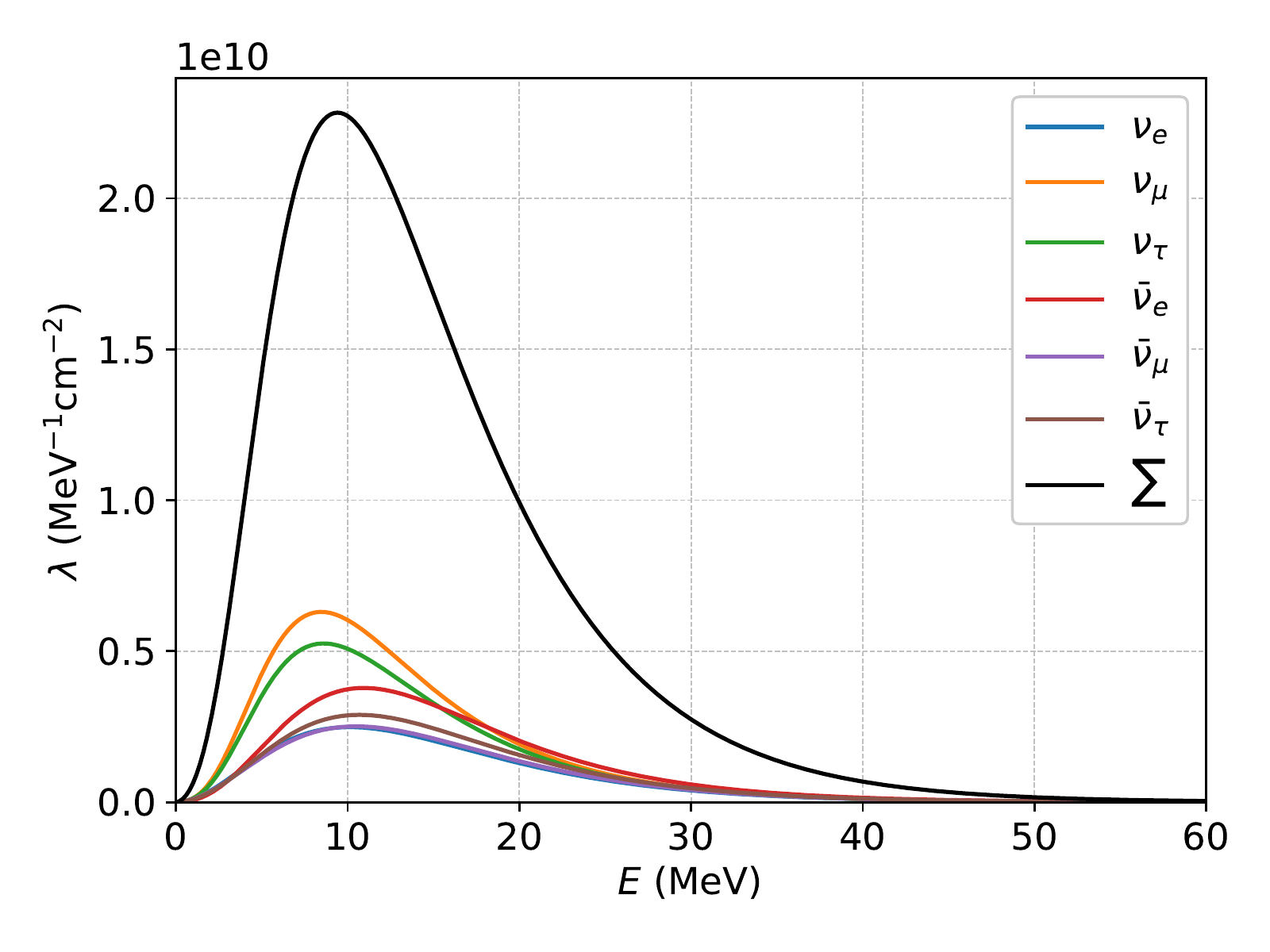}{0.5\textwidth}{(b)}
    \fig{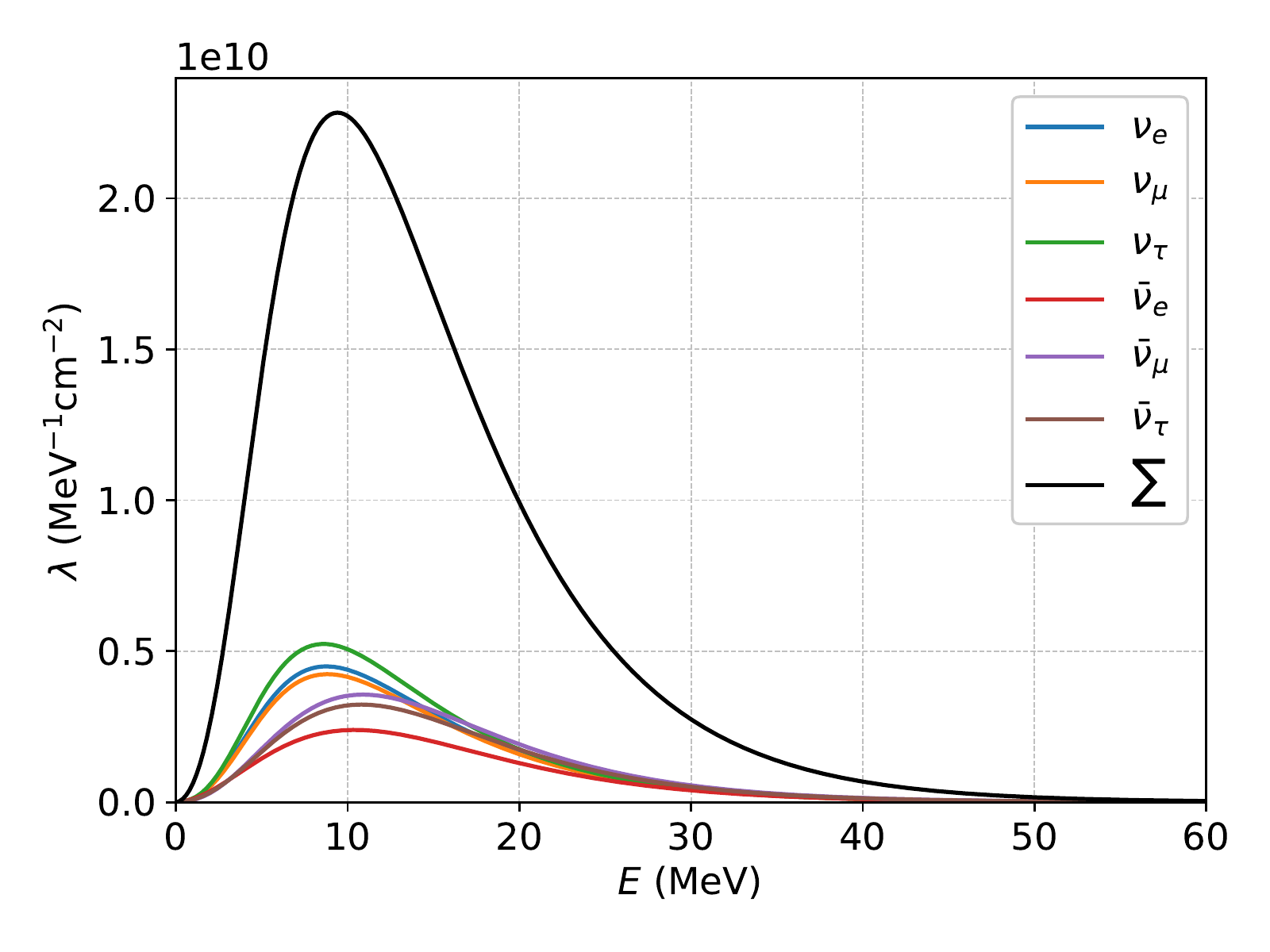}{0.5\textwidth}{(c)}
    }
	\caption{Fluences corresponding to the SN model LS220-s15.0 \citep{web:Garching}, assuming a distance to the SN equal to 10 kpc, considering (a) no neutrino flavor transformation in the SN matter and the adiabatic MSW effect with (b) NO and (c) IO. The full SN simulation time interval is considered. In the case of NO, fluences corresponding to $\nu_e$ and $\bar{\nu}_{\mu}$ are superimposed.} 
	\label{fig:Fluences}  
\end{figure}

The EstrellaNueva software was specifically developed for complementing the simulations made by various supernova simulation groups in estimating the interaction rates expected in detectors. Currently, there are implemented several supernova models in the software package. The models are the result of simulations performed by the Core-Collapse Modeling Group at the Max Planck Institute for Astrophysics~\citep{web:Garching}, that reports the time dependence of the primary flux parameters at the neutrino-sphere, namely $\mathcal{L}_{\nu}(t)$, $\langle E_{\nu} \rangle(t)$, and $\langle E^2_{\nu} \rangle(t)$. The time dependence of $\beta_{\nu}$ is calculated by EstrellaNueva following Equation (\ref{eq:beta_en_t}). These models were simulated with the PROMETEUS-VERTEX code \citep{M_ller_2010,PhysRevD.85.085031} using three different equations of state: the equation of state (EoS) of Lattimer and Swesty \citep{LATTIMER1991331}, with a nuclear incompressibility of 220 MeV, denoted by LS220; the Shen EoS \citep{SHEN1998435}; and the SFHo hadronic EoS \citep{Steiner_2013}. Implemented SN models are described in Table \ref{tab:models}, where the time intervals in which they were simulated are also shown. As an example, Figure \ref{fig:SN_model} shows the time dependence of the primary flux parameters for the model LS220-s15.0, which corresponds to a progenitor of mass 15.0 $\text{M}_{\odot}$ ($\text{M}_{\odot}$ is the mass of the Sun) and was simulated using the LS220 EoS. Figure \ref{fig:Fluences} shows the fluences for this model at a distance to the SN equal to 10 kpc, considering the MSW effect with NO, IO, and neglecting the neutrino flavor transformation in the SN matter. For fluence calculations, the entire time interval of the model was used.

\begin{table}[]
    \centering
    \begin{tabular}{l|c|c|c|c}
         \hline\hline
         Model&Progenitor Mass [M${}_\odot$] &EoS &Time interval [s]&Reference\\
         \hline
         LS220-s11.2&11.2&LS220&-0.170092974551 -- 0.500004733154&\citep{web:Garching}\\
         LS220-s11.2c&11.2&LS220&-0.170092974551 -- 7.600478503686&\citep{Mirizzi:2015eza,web:Garching}\\
         LS220-s12.0&12.0&LS220&-0.187296800690 -- 0.497543448537&\citep{web:Garching}\\
         LS220-s15.0&15.0&LS220&-0.304032523673 -- 0.498446483650&\citep{web:Garching}\\
         LS220-s15s7b2&15.0&LS220&-0.221282374565 -- 0.496251020000&\citep{web:Garching}\\
         LS220-s17.6&17.6&LS220&-0.284925820013 -- 0.498178366419&\citep{web:Garching}\\
         LS220-s17.8&17.8&LS220&-0.279131376354 -- 0.500068432690&\citep{web:Garching}\\
         LS220-s20.0&20.0&LS220&-0.256519704513 -- 0.499980285627&\citep{web:Garching}\\
         LS220-s20.6&20.6&LS220&-0.412889509135 -- 0.500009131565&\citep{web:Garching}\\
         LS220-s25.0&25.0&LS220&-0.455379448410 -- 0.499728493980&\citep{web:Garching}\\
         LS220-s27.0&27.0&LS220&-0.344627618270 -- 0.497878869527&\citep{web:Garching}\\
         LS220-s27.0c&27.0&LS220&-0.344627618270 -- 8.350202515285&\citep{Mirizzi:2015eza, web:Garching}\\
         LS220-s27.0co&27.0&LS220&-0.349455361517 -- 15.439433579436&\citep{Mirizzi:2015eza,web:Garching}\\
         LS220-z9.6co&9.6&LS220&-0.233381022196 -- 11.999932339246&\citep{ Mirizzi:2015eza,web:Garching}\\   LS220-s40.0c&40.0&LS220&-0.408731709799 -- 2.105601215751&\citep{Mirizzi:2015eza,web:Garching}\\
         LS220-s40s7b2c&40.0&LS220&-0.408737176878 -- 0.567932280255&\citep{Mirizzi:2015eza,web:Garching}\\
         SFHo-s27.0co&27.0&SFHo&-0.292910194783 -- 11.169139356183&\citep{web:Garching}\\
         SFHo-z9.6co&9.6&SFHo&-0.242266894100 -- 13.622987243331&\citep{ Mirizzi:2015eza,web:Garching}\\
         Shen-s8.8&8.8&Shen&-0.0636472236 -- 8.90975001&\citep{ PhysRevLett.104.251101,web:Garching}\\
         Shen-s11.2&11.2&Shen&-0.136560972782 -- 0.495312224658&\citep{web:Garching}\\
         Shen-s12.0&12.0&Shen&-0.144379115573 -- 0.495545475007&\citep{web:Garching}\\
         Shen-s15.0&15.0&Shen&-0.211057570069 -- 0.499739838584&\citep{web:Garching}\\
         Shen-s15s7b2&15.0&Shen&-0.167861670324 -- 0.500369261592&\citep{web:Garching}\\
         Shen-s17.6&17.6&Shen&-0.199208094334 -- 0.498790872514&\citep{web:Garching}\\
         Shen-s17.8&17.8&Shen&-0.190373490538 -- 0.498689914046&\citep{web:Garching}\\
         Shen-s20.0&20.0&Shen&-0.187767329639 -- 0.500071977502&\citep{web:Garching}\\
         Shen-s20.6&20.6&Shen&-0.254702334699 -- 0.499291328668&\citep{web:Garching}\\
         Shen-s25.0&25.0&Shen&-0.275634063374 -- 0.496790109348&\citep{web:Garching}\\
         Shen-s27.0&27.0&Shen&-0.227437666075 -- 0.496242346477&\citep{web:Garching}\\
         Shen-s40.0&40.0&Shen&-0.254058543569 -- 0.500396111047&\citep{web:Garching}\\
         \hline\hline
    \end{tabular}
    \caption{Supernova models from the Core-Collapse Modeling Group at the Max Planck Institute for Astrophysics~\citep{web:Garching}, available in the software EstrellaNueva.}
    \label{tab:models}
\end{table}

In addition, it is possible to define a simple and time-independent SN model in the EstrellaNueva software. This is assuming the primary flux parameters constant in time. In this scenario, the fluences of the primary fluxes are given by the following expression:
\begin{equation}
    \begin{split}
        \lambda_{\nu}^0 &= \frac{1}{4\pi d^2}\int_{t_0}^{t_{end}}F_{\nu}^0(E,t)dt\\
        &=\frac{\varepsilon_{\nu}}{4\pi d^2} \frac{\left(1+\beta_{\nu}\right)^{1+\beta_{\nu}}}{\Gamma\left(1+\beta_{\nu}\right)}\frac{E^{\beta_{\nu}}}{\left(\left<E_{\nu}\right>\right)^{\beta_{\nu}+2}}\times\text{exp}\left[-\left(\beta_{\nu}+1\right)\frac{E}{\left<E_{\nu}\right>}\right],
    \end{split}
\label{eq:non_t_dependent_fluence}
\end{equation}
where $\varepsilon_{\nu}$ is the total energy emitted from the neutrino-sphere in form of neutrinos of the specie $\nu$. Then, if the adiabatic MSW effect is considered, the flavor transformed fluences $\lambda_{\nu}$ are linear combinations of the primary fluences $\lambda_{\nu}^0$, which are calculated following Equations (\ref{eq:oscillated_fluxes_NO}) or (\ref{eq:oscillated_fluxes_IO}) depending on the neutrino mass ordering assumed. The primary fluxes $F_{\nu}^0$ are replaced by the primary fluences $\lambda_{\nu}^0$ for this case. The values of the parameters $\varepsilon_{\nu}$, $\left<E_{\nu}\right>$, and $\beta_{\nu}$ are requested as input for the neutrino species considered in the neutrino-sphere. 

\section{Detection channels}\label{sec:detection_channels}

EstrellaNueva has implemented the cross sections of the principal detection channels for materials employed by current and future detectors, such as water, argon, xenon, lead, and mineral liquid scintillators. In addition, the software provides the flexibility to define materials with any chemical composition. The detection channels implemented are: $\nu-e$ elastic scattering, $\nu-p$ elastic-scattering, inverse beta decay (IBD), coherent elastic neutrino-nucleus scattering (CE$\nu$NS), and the interactions with ${}^{12}$C, ${}^{16}$O, and ${}^{208}$Pb. The last three were taken from the SNOwGLoBES repository \citep{snowglobes} in the form of data files and the rest were implemented analytically. The implementation of analytical cross sections is one of the main features provided by this software package.

\begin{figure}[h!]
	\gridline{\fig{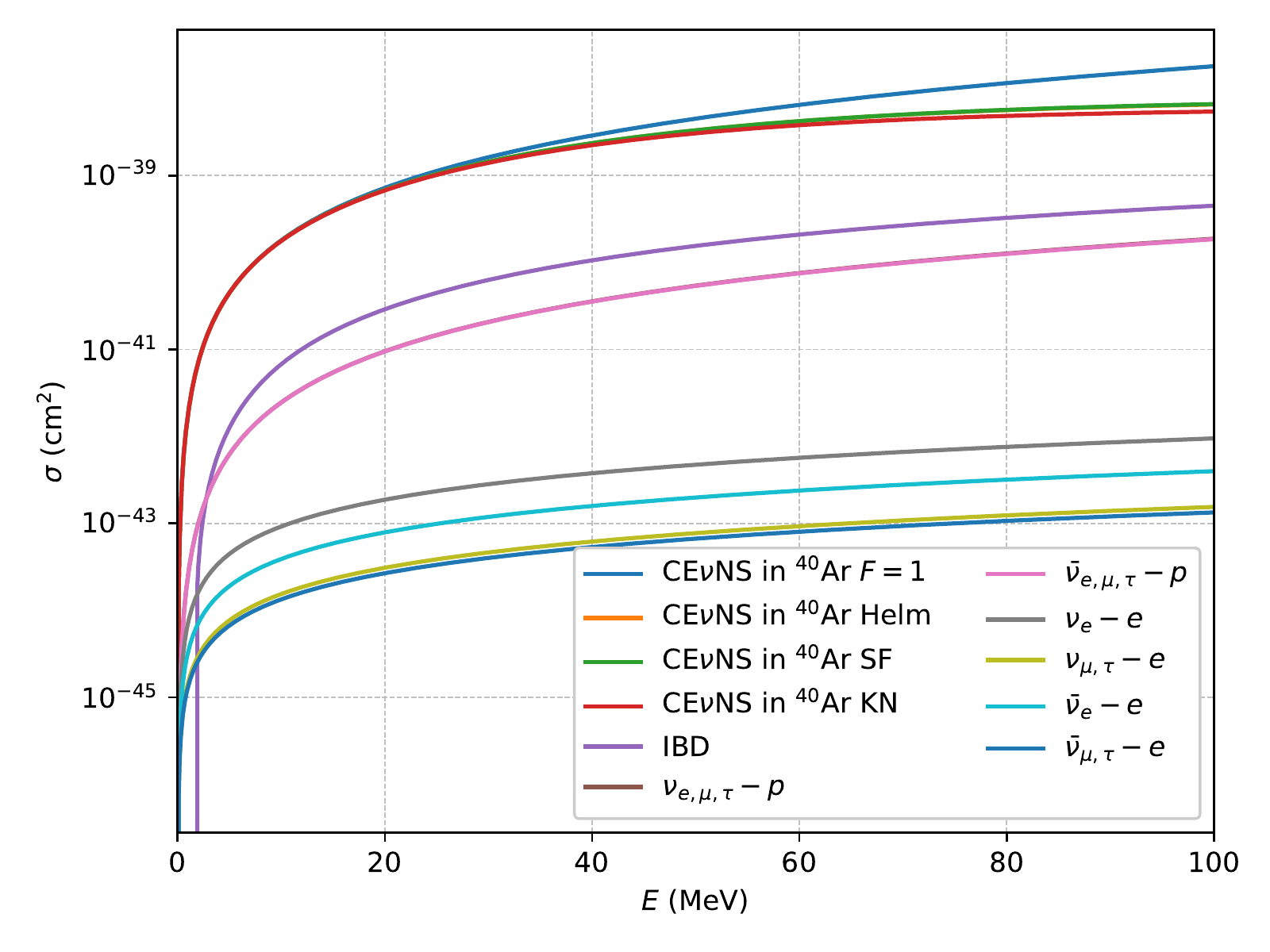}{0.7\textwidth}{(a)}}
	\gridline{\fig{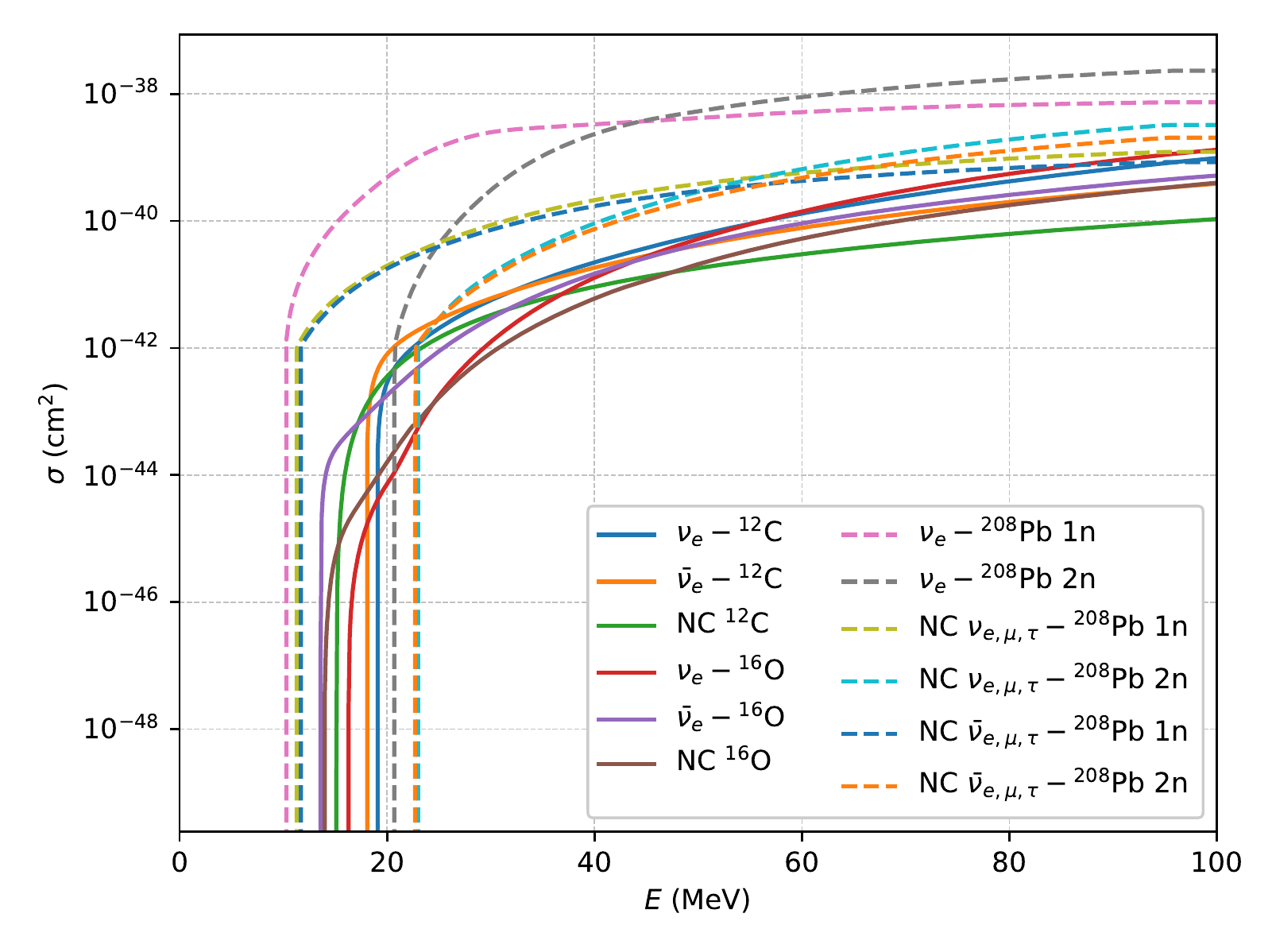}{0.7\textwidth}{(b)}
    }
	\caption{Cross sections as function of the incoming neutrino energy $E$. (a) In the case of scattering interactions with protons, the cross sections for neutrinos and antineutrinos are superimposed. The cross sections for CE$\nu$NS in ${}^{40}$Ar corresponding to the Helm and SF form factors are also superimposed. $F=1$ assumes a form factor equal to one. (b) NC are neutral current interactions, while the rest are charge current interactions. In the case of interactions with ${}^{208}\text{Pb}$, $\left(\text{1n}\right)$ and $\left(\text{2n}\right)$ result in one and two neutrons, respectively. See \citep{snowglobes} for references.}
	\label{fig:cross_section}  
\end{figure}

The analytical expression for the $\nu-e$ elastic scattering differential cross section, taken from \citep{Giunti:2014ixa}, is
\begin{equation}
\frac{d\sigma_{\nu}}{dT}(E,T)=\frac{G_F^2m_e}{2\pi}\left\{\left(g_V^{\nu}+g_A^{\nu}\right)^2+\left(g_V^{\nu}-g_A^{\nu}\right)^2\left(1-\frac{T}{E}\right)^2+\left[\left(g_A^{\nu}\right)^2-\left(g_V^{\nu}\right)^2\right]\frac{m_eT}{E^2}\right\},
\label{eq:cross_section_standartd} 
\end{equation}
where $m_e$ is the electron mass and $G_F$ the Fermi coupling constant. The coupling constants $g_V^{\nu}$ and $g_A^{\nu}$ are related to the effective weak mixing angle $\theta_{W}$ ($\sin^2 \theta_{W}=0.23153$ \citep{RevModPhys.93.025010}) as follows, 
 \begin{align}
&g_V^{\nu_e}=2\sin^2\theta_W + 1/2, \hspace{1cm} g_A^{\nu_e}=1/2,\\
&g_V^{\nu_{\mu,\tau}}=2\sin^2\theta_W-1/2,\hspace{.75cm} g_A^{\nu_{\mu,\tau}}=-1/2,\\
&g_V^{\bar{\nu}_e}=2\sin^2\theta_W + 1/2, \hspace{1cm} g_A^{\bar{\nu}_e}=-1/2,\\
&g_V^{\bar{\nu}_{\mu,\tau}}=2\sin^2\theta_W-1/2,\hspace{.75cm} g_A^{\bar{\nu}_{\mu,\tau}}=1/2.
\end{align}

\color{black}For $\nu-p$ elastic scattering, the differential cross section, taken from \citep{PhysRevD.66.033001}, is given by 
\begin{equation}
    \frac{d\sigma_{\nu}}{dT}(E,T) = \frac{G_F^2 m_p}{2\pi E^2}\left[\left(c_V\pm c_A\right)^2 E^2 + \left(c_V\mp c_A\right)^2(E-T)^2 - (c_V^2 - c_A^2)m_p T\right],
\end{equation}
where $m_p$ represents the proton mass. $c_A$ and $c_V$ are the axial-vector and vector coupling constants between the exchanged $Z^0$ boson and the proton, with 
\begin{equation}
    c_V = \frac{1-4\sin^2 \theta_W}{2} \hspace{1cm} \text{and}\hspace{1cm} c_A = \frac{g_A(0)\cdot(1 + \eta)}{2},
\end{equation}
where $g_A(0)=1.267$ is the axial proton form factor \citep{PhysRevD.86.010001} and $\eta= 0.12$ is the proton  strangeness \citep{PhysRevD.35.785}. The upper sign refers to neutrinos and the lower to anti-neutrinos.\\

For the IBD interaction a simple but accurate approximation for SN neutrino energies ($E<300$ MeV) is implemented. The total cross section as a function of the incoming $\bar{\nu}_e$ energy is given by \citep{STRUMIA200342} 
\begin{equation}
    \sigma_{\text{IBD}}(E) = 10^{-43}\text{cm}^2 p_{e^{+}} E^{\text{tot}}_{e^{+}}E^{- 0.07056+0.02018\ln E - 0.001953\ln^3E},  
\end{equation}
where $E_{e^+}^{\text{tot}}\approx E - (m_{n} - m_{p})$, with $m_{n}$ representing the neutron mass; and $p_{e^+}$ is the positron momentum. This interaction has a threshold energy given by 
\begin{equation}
    E_{\text{thr}} = \frac{(m_n + m_e)^2 - m_p}{2m_p}\approx 1.806 \text{ MeV}. 
\end{equation}

Lastly, in the case of CE$\nu$NS, the differential cross section is taken from \citep{CEnuNS_cross_section} and given by the following expression, 
\begin{equation}
    \frac{d\sigma}{dT}(E,T) = \frac{G_F^2 M}{4\pi}q_W^2\left(1 - \frac{MT}{E^2}\right)F^2(q^2),
\end{equation}
where $M$ is the target nucleus mass,  $q_W = N - \left(1 - 4\sin^2 \theta_W\right)Z$ is the weak nuclear charge with $N$ and $Z$ the number of neutrons and protons in the target nucleus, respectively. $F(q^2)$ is the nuclear form factor as function of the transferred momentum $q$. The following nuclear form factors are implemented: the Helm form factor, the symmetrized Fermi (SF) form factor, and the Klein-Nystrand (KN) form factor. A nuclear form factor equal to one can also be assumed. Next, the analytical expressions of the nuclear form factors will be discussed.

The analytical expression for the Helm form factor parameterization is given by \citep{LEWIN199687, PAPOULIAS2020135133} 
\begin{equation}
F_{\text{Helm}}(q^2) = 3\frac{j_1(q R_0)}{q R_0}e^{-(q s)^2/2},    
\label{eq:Helm_form_factor}
\end{equation} 
where $j_1$ denotes the 1st-order spherical Bessel function, $s=0.9$ fm is the  nuclear surface thickness, and $R_0 = \sqrt{c^2 + \frac{7}{3}\pi^2a^2 + s^2}$ is the effective nuclear radius, the so-called diffraction radius. $c = 1.23A^{1/3} - 0.60$ fm and $a = 0.52$ fm represent the half density radius and the diffuseness, respectively, where $A$ is the mass number of the target nucleus.

The analytical expression for the SF form factor is given by \citep{PAPOULIAS2020135133}
 \begin{equation}
     F_{\text{SF}}(q^2) = \frac{3}{q c\left[(q c)^2 + (\pi q a)^2\right]}\left[\frac{\pi q a}{\sinh{(\pi q a)}}\right]\left[\frac{\pi q a \sin{(q c)}}{\tanh{(\pi q a)}} - q c \cos{(q c)}\right]. 
 \end{equation}
 
 Finally, in the case of the KN form factor, the implemented analytical expression is given by \citep{ PAPOULIAS2020135133, CEnuNS_cross_section}
 \begin{equation}
     F_{\text{KN}} = \frac{4\pi\rho_0}{Aq^3}(\sin(qR_A) - qR_A\cos(qR_A))\frac{1}{1 + (q a_k)^2},
 \end{equation}
where $a_k = 0.7$ fm is the range of a Yukawa potential which is convoluted over a Woods-Saxon distribution, approximated as a hard sphere with radius $R_A = A^{1/3}r_0$, with the proton radius $r_0 = 1.3$ fm, and $\rho_0 = \frac{3}{4\pi r_0^3}$ is the nuclear density.  $q^2\approx 2MT$ is considered to evaluate the nuclear form factors as functions of the recoil nucleus energy.

\begin{figure}
    \centering
    \includegraphics[width=.7\linewidth]{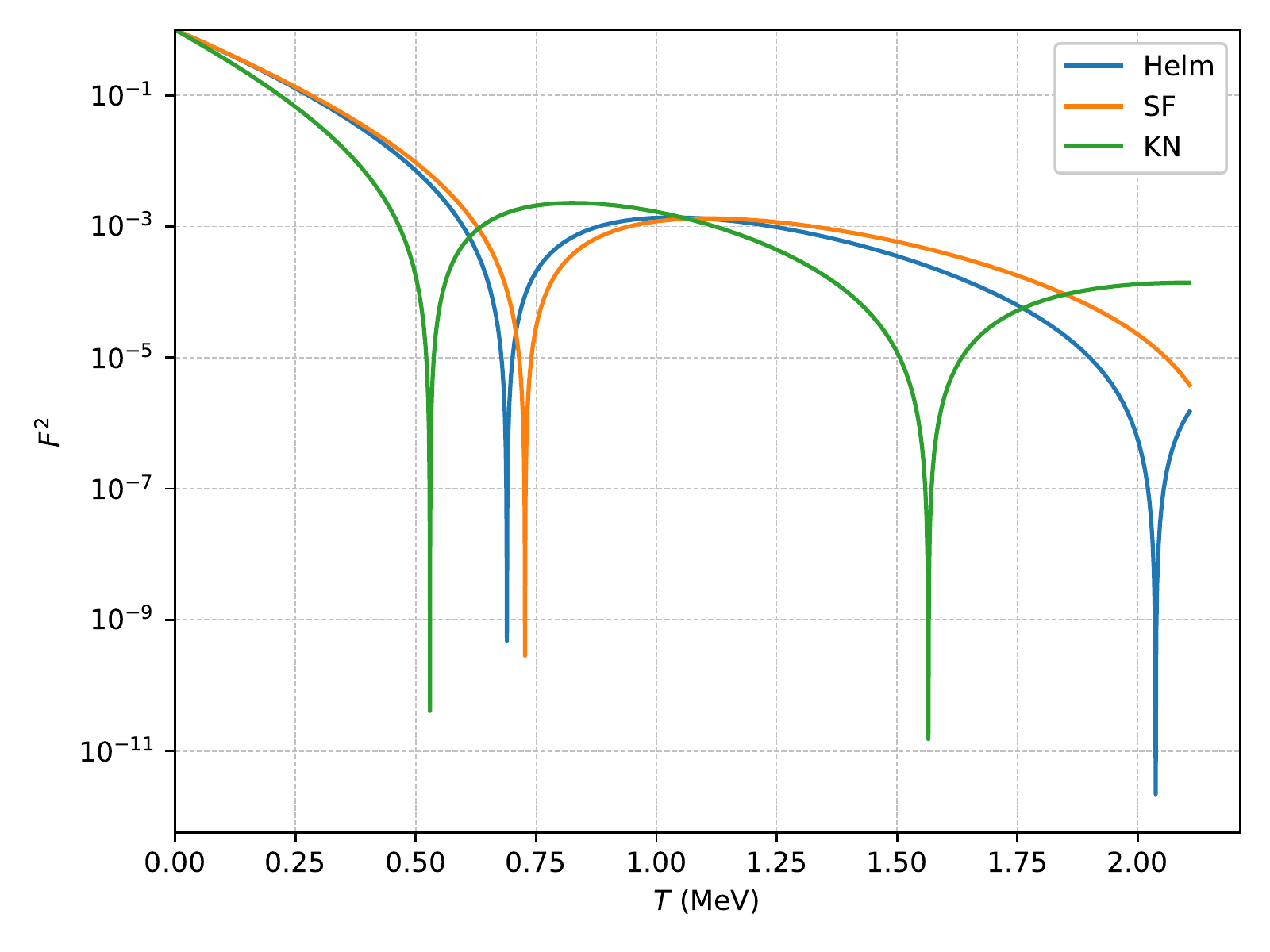}
    \caption{Nuclear form factors for ${}^{40}$Ar as function of the nucleus recoil energy $T$.}
    \label{fig:form_factors}
\end{figure}

Figure \ref{fig:cross_section} shows the cross sections implemented in EstrellaNueva, with the cross section for CE$\nu$NS in ${}^{40}$Ar (as an example) using the form factors considered in the software. Figure \ref{fig:form_factors} shows the Helm, SF, and KN form factors for ${}^{40}$Ar.

\section{Configuration file examples for EstrellaNueva}
\label{sec:usage_examples}

The EstrellaNueva software is a standalone code that does not require any more dependencies other than the Python
modules that were used for its programming. The modules are Numpy \citep{Numpy}, Scipy \citep{Scipy}, and Mathplotlib \citep{Hunter:2007}. These modules are required in the Python3 environment.  %The GitHub repository is located at \url{https://github.com/snoplus/EstrellaNueva.git}.

% In order to run the code, the user needs to configure the execution via the configuration file ``\texttt{config.json}", located in the main directory ``EstrellaNueva". Once the configuration was performed the user has to run the main file “EstrellaNueva.py” with the interpreter “python3” from the “EstrellaNueva” directory. The command line is: ``\texttt{python3 EstrellaNueva.py}".  

In this section, two examples of the configuration file are presented to calculate interaction rates. The SN model LS220-s15.0, presented in Sec. \ref{sec:supernova_models}, is used as example, with the distance to the SN assumed at 10 kpc and considering the full SN simulation time interval. The fluences regarding this configuration for the neutrino flavor transformation scenarios implemented (adiabatic MSW effect with normal and inverted neutrino mass ordering, and no neutrino flavor transformation) are shown in Figure \ref{fig:Fluences}.\\

\subsection{Interaction rates for $\nu-e$ elastic scattering in a 100 kton detector using liquid scintillator}

\begin{figure}[htbp]
	\gridline{\fig{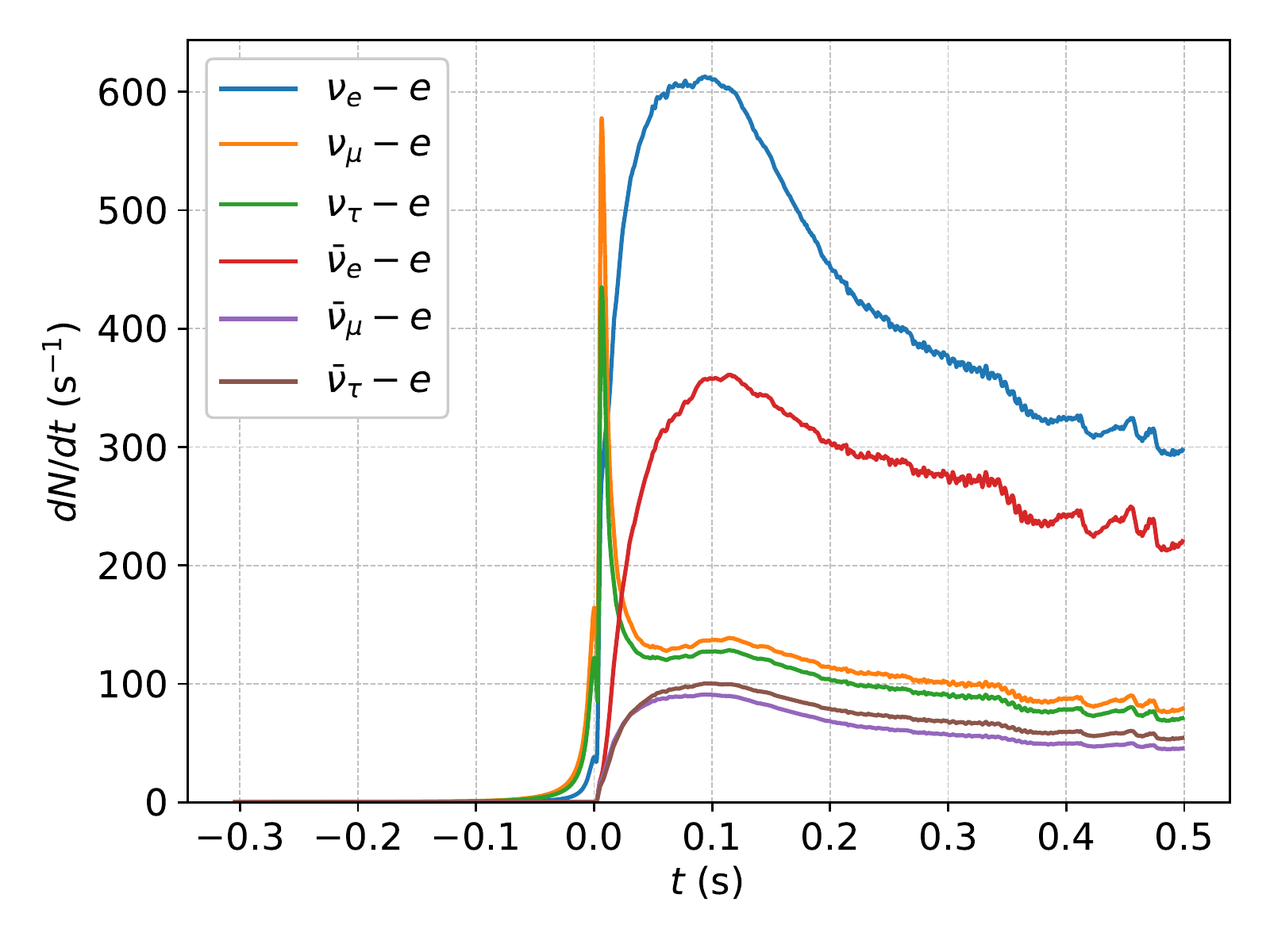}{0.5\textwidth}{(a)}
    }
    \gridline{\fig{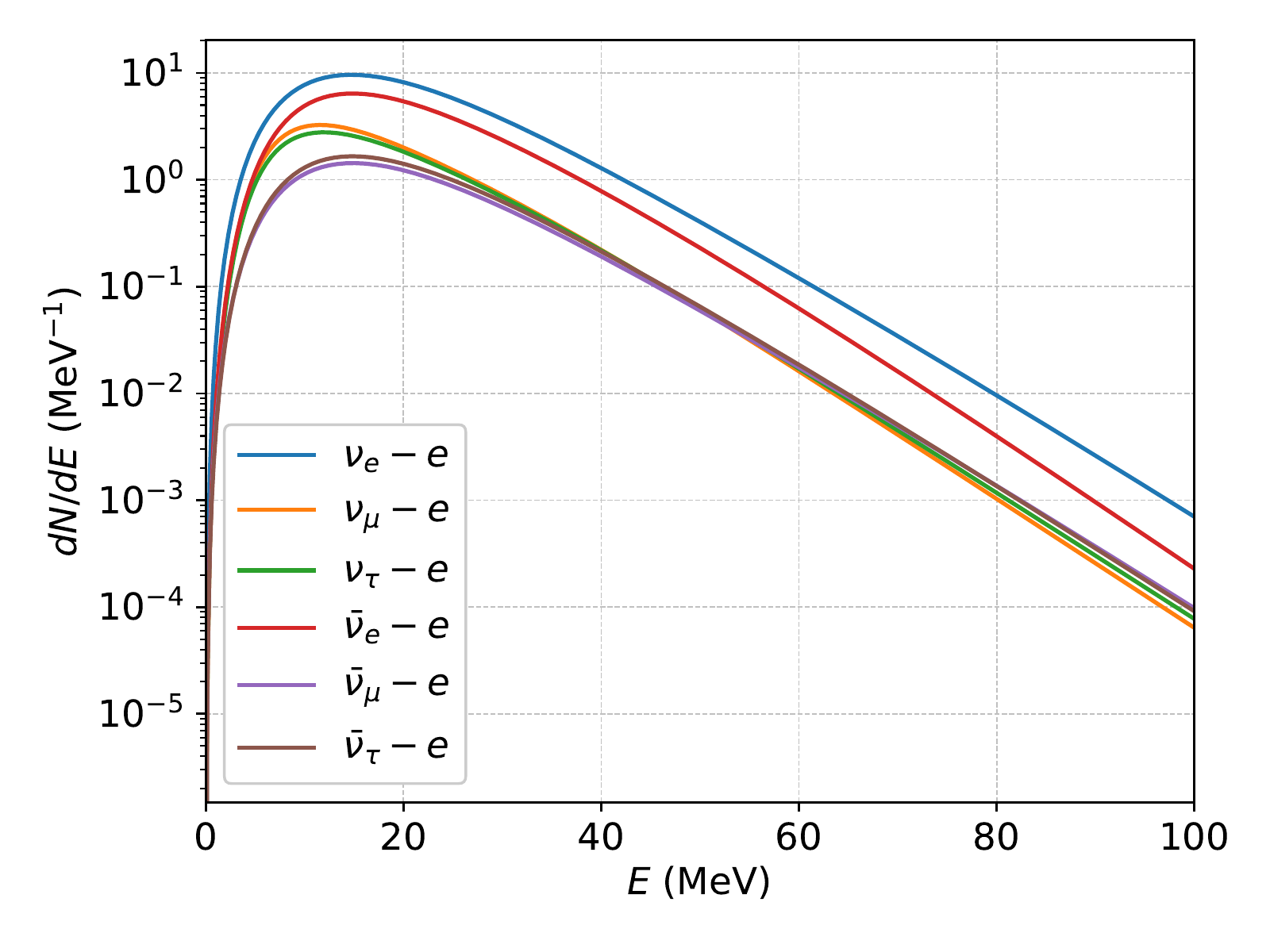}{0.5\textwidth}{(b)}
    \fig{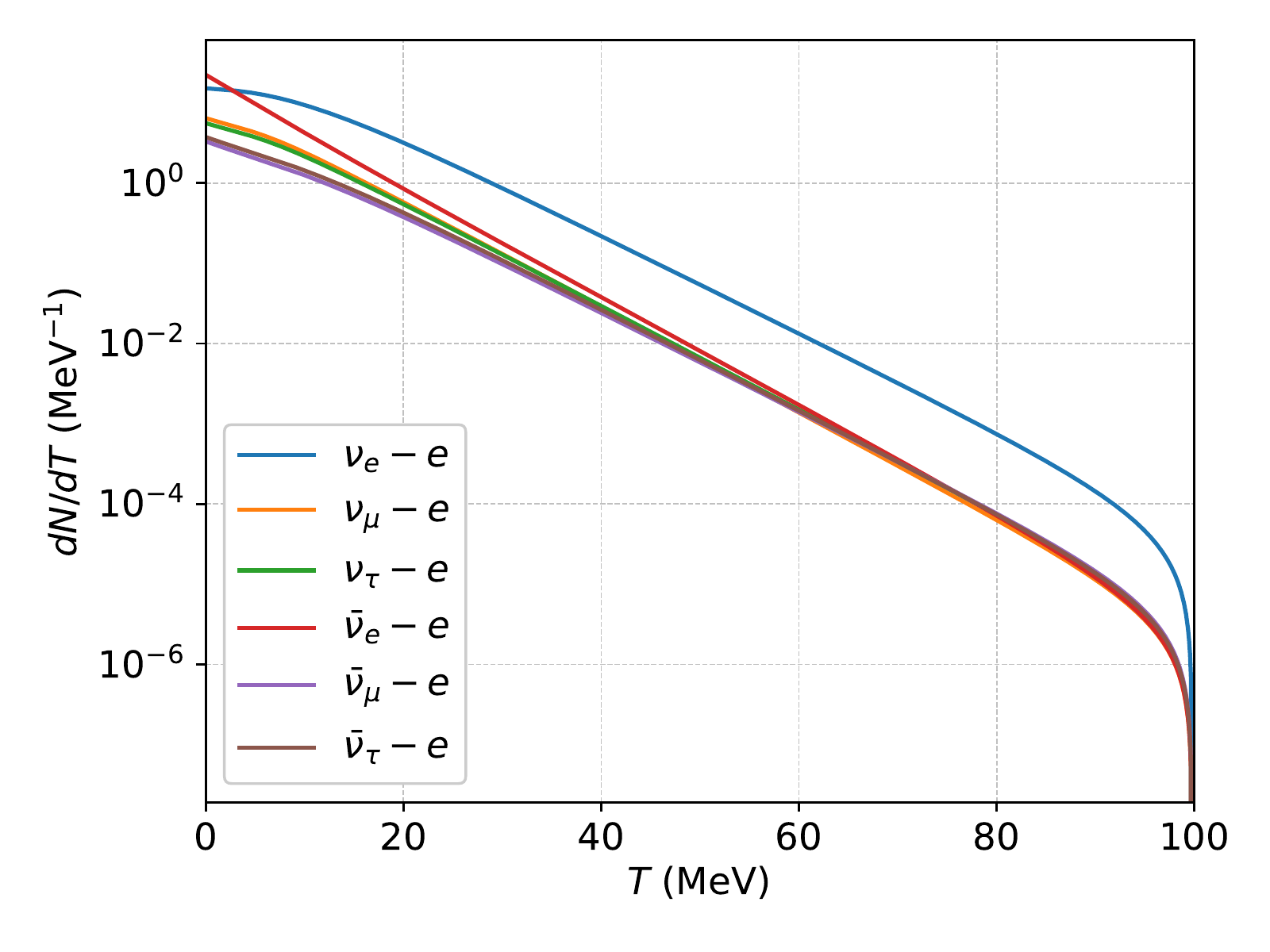}{0.5\textwidth}{(c)}
    }
	\caption{Interaction rates with respect to (a) time $t$, (b) incoming neutrino energy $E$, and (c) recoil electron energy $T$ for $\nu-e$ elastic scattering in 100 kton of LAB. The NO scenario for the adiabatic MSW flavor transformation is assumed and model LS220-s15.0 is used at a distance of 10 kpc. The full SN simulation time interval is considered.}
	\label{fig:event_rates_nu_e}  
\end{figure}

In this example, the interaction rates for $\nu-e$ elastic scattering are calculated for a 100 kton generic detector with linear alkylbenzene (LAB). LAB is used as a target medium by the SNO+ experiment \citep{Albanese_2021} and it will be used by the future JUNO experiment \citep{CAO2019230}. LAB is a material internally predefined in the software, although any chemical composition can be defined in the configuration file. The configuration to compute the interaction rates, considering normal neutrino mass ordering (NO), is as follows, 
\begin{verbatim}
{
"model": "ls220_s15.0",
"msw": "NO",
"distance": 10,
"material": "LAB",
"detector_mass": 100,
"interactions": ["nue_e", "numu_e", "nutau_e", "nuebar_e", "numubar_e", "nutaubar_e"]
}
\end{verbatim}

Figure \ref{fig:event_rates_nu_e} shows the interaction rates obtained from the above configuration. The total number of events for each channel are shown in Table \ref{tab:number_of_events}. %In order to compute the interaction rates for IO or without considering neutrino oscillations, the user must set the \texttt{"IO"} or \texttt{"off"} values in the \texttt{"msw"} parameter respectively.

\begin{table}[h]
    \centering
    \begin{tabular}{c|r}
        \hline \hline
        Channel & Events  \\
        \hline
        $\nu_e-e$ &212.21\\
        $\nu_{\mu}-e$&59.94\\
        $\nu_{\tau}-e$&53.30\\
        $\bar{\nu}_e-e$&135.64\\
        $\bar{\nu}_{\mu}-e$&31.53\\
        $\bar{\nu}_{\tau}-e$&35.93\\
        \hline
        Total&528.55\\
        \hline\hline
    \end{tabular}
    \caption{Number of interactions for $\nu-e$ elastic scattering in 100 kton of LAB, in the NO scenario for the MSW flavor transformation, and assuming model LS220-s15.0 at a distance of 10 kpc. The full SN simulation time interval is considered.}
    \label{tab:number_of_events}
\end{table}

\subsection{Interaction rates for CE$\nu$NS in a 100 tonne detector using liquid argon}

\begin{table}[h!]
    \centering
    \begin{tabular}{c|c}
        \hline \hline
        Nuclear form factor & Events  \\
        \hline
        $F$ = 1 &279.13\\
        Helm&255.11\\
        SF&255.17\\
        KN&249.34\\
        \hline\hline
    \end{tabular}
    \caption{Number of interactions for CE$\nu$NS in 100 tonnes of ${}^{40}$Ar for the form factors considered. The SN model LS220-s15.0 is used assuming a distance to the SN of 10 kpc and considering no neutrino flavor transformation in the SN matter. The full SN simulation time interval is considered.}
    \label{tab:number_of_events_CEnuNS}
\end{table}

\begin{figure}[htbp]
	\gridline{\fig{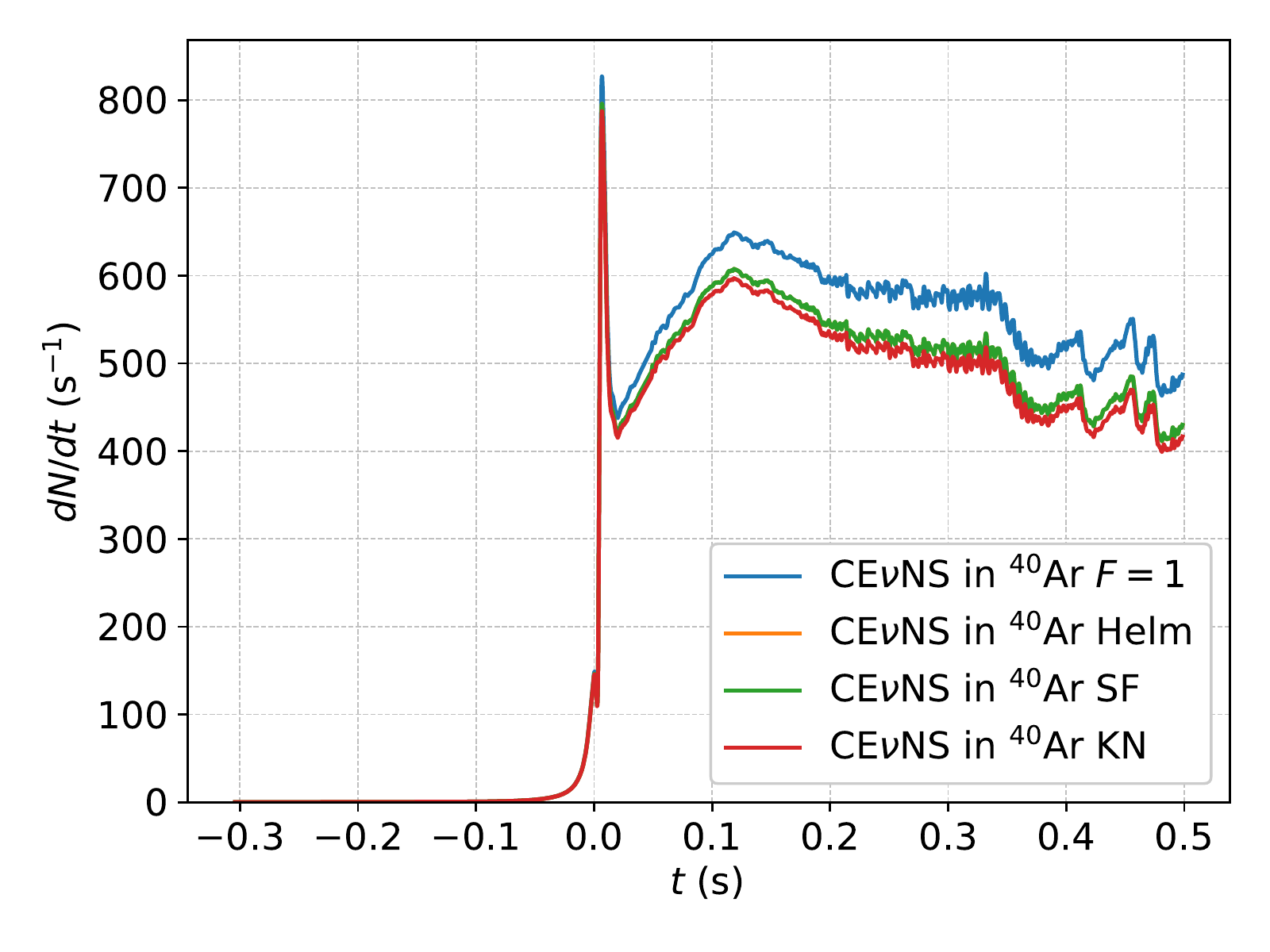}{0.5\textwidth}{(a)}
    }
    \gridline{\fig{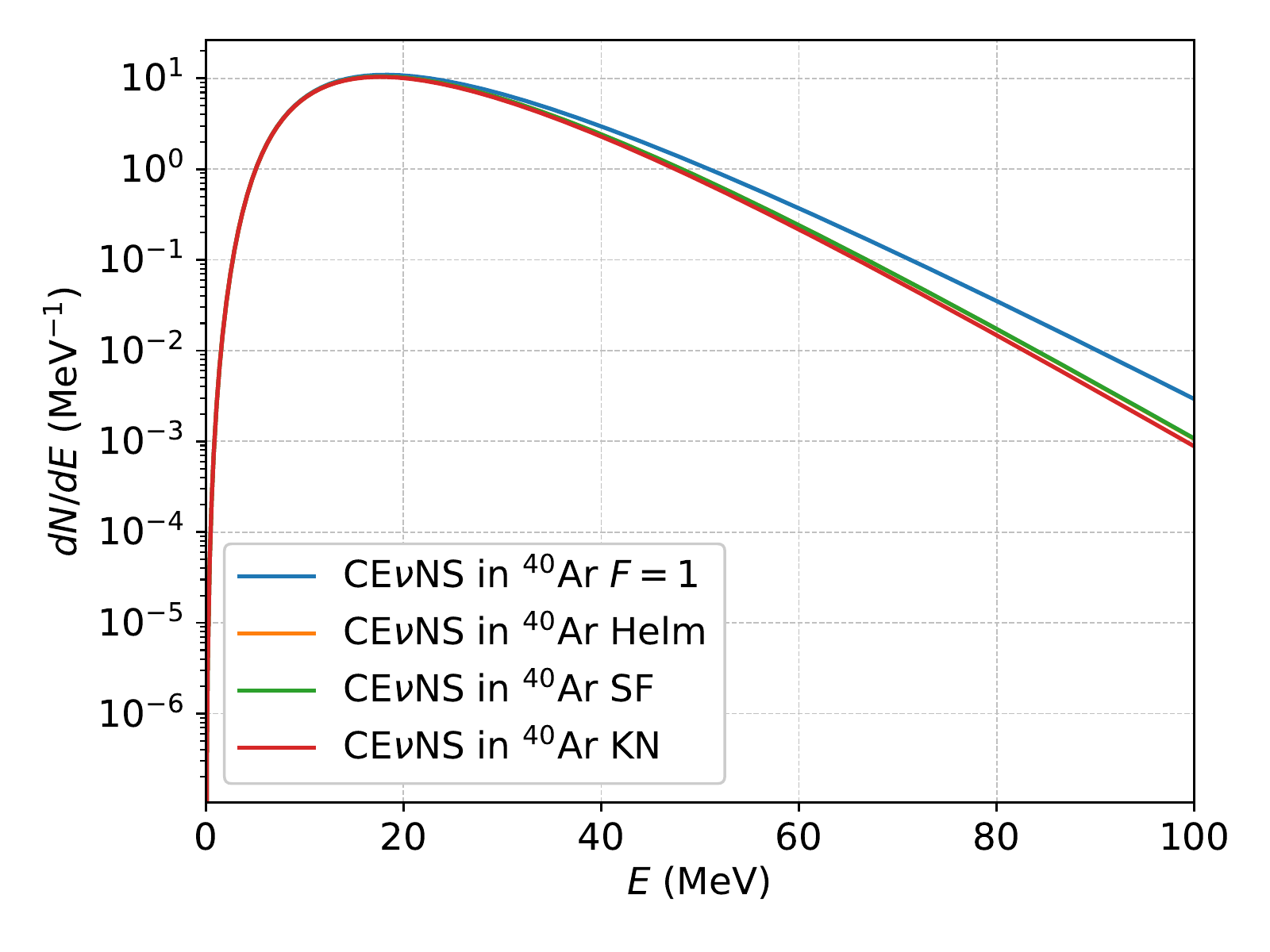}{0.5\textwidth}{(b)}
    \fig{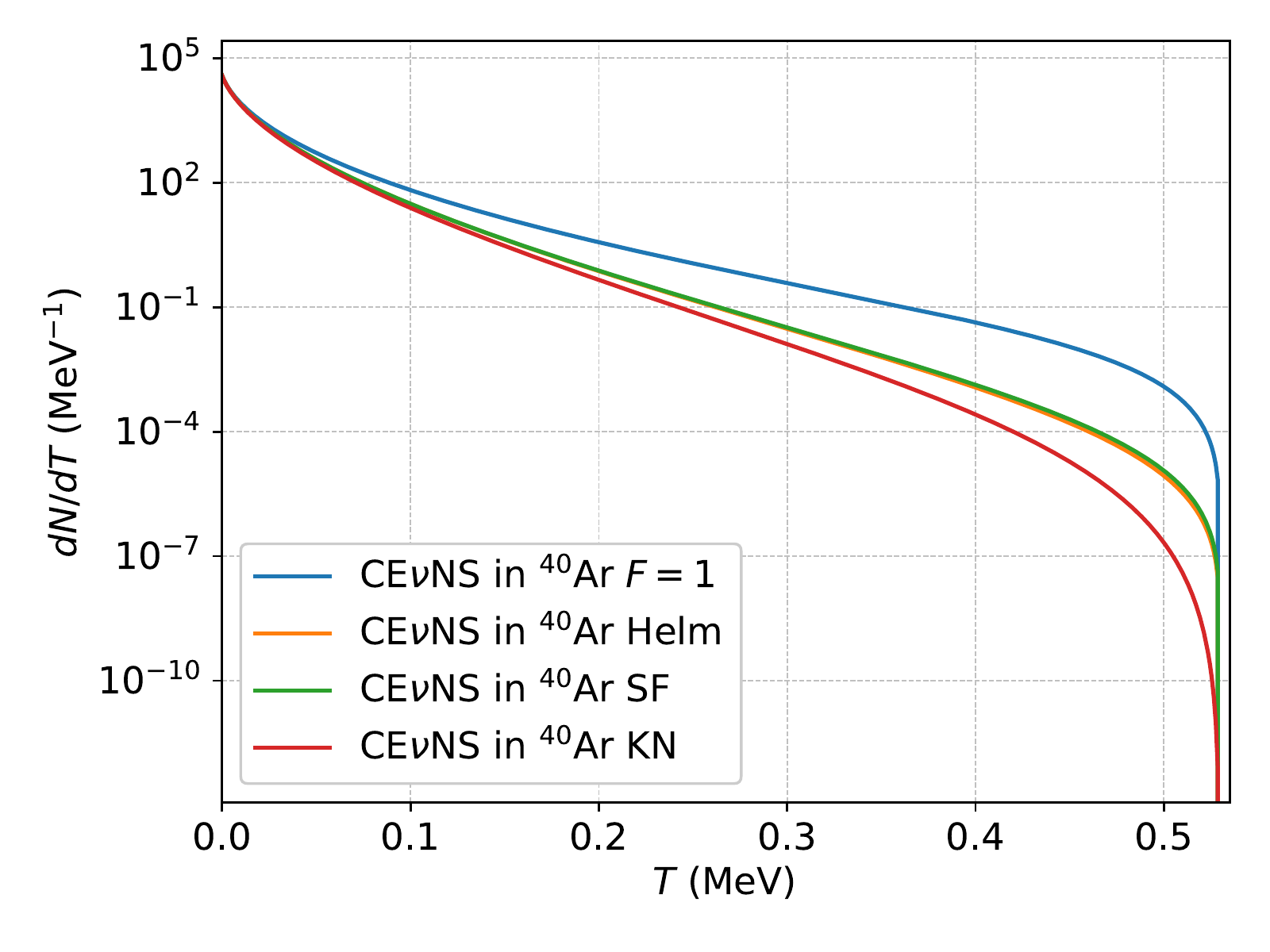}{0.5\textwidth}{(c)}
    }
	\caption{Interaction rates for CE$\nu$NS in 100 tonnes of ${}^{40}$Ar. The SN model LS220-s15.0 is used assuming a distance to the SN of 10 kpc and considering no neutrino flavor transformation in the SN matter. The full SN simulation time interval is considered. The interaction rates for the Helm and the SF nuclear form factors are superimposed. Rates are shown with respect to (a) time $t$, (b) incoming neutrino energy $E$, and (c) recoil nucleus energy $T$.} 
	\label{fig:event_rates_CEnuNS}  
\end{figure}

This example corresponds to the event rates calculated for CE$\nu$NS in 100 tonnes of liquid argon, considering a chemical composition of pure ${}^{40}$Ar. Argon is also predefined internally. CE$\nu$NS interactions are not affected by flavor transformations and they are turned off in this case. The results do not change if flavor transformations are included. The configuration file for the implemented nuclear form factors is as follows:  

\begin{verbatim}
{
"model": "ls220_s15.0",
"msw": "off",
"distance": 10,
"material": "Argon",
"detector_mass": 0.1,
"interactions": ["CEnuNS_Ar40_one", "CEnuNS_Ar40_Helm", "CEnuNS_Ar40_Symmetrized_Fermi",
"CEnuNS_Ar40_Klein_Nystrand"]
}
\end{verbatim}

The interaction rates obtained from the above configuration file are shown in Figure \ref{fig:event_rates_CEnuNS}. Table \ref{tab:number_of_events_CEnuNS} shows the number of interactions for each nuclear form factor. A complete and detailed guide of the EstrellaNueva software is included in the Zenodo repository.

\section{Conclusions}
\label{sec:conclusions}

Several detectors all over the world are currently expecting the emergence of a supernova. Thousands of events are expected in current operating neutrino detectors. Simple, flexible, standalone, and independent tools are valuable and will be useful when such a cosmic event takes place. The EstrellaNueva software provides a user-friendly tool to study neutrinos and supernova properties by calculating the number of events in detectors. Several detection channels are currently implemented, the majority with analytical functions, such as neutrino-electron and neutrino-proton elastic scattering, inverse beta decay, and coherent elastic neutrino-nucleus scattering with the Helm, Klein-Nystrand, and symmetrized Fermi nuclear form factors. The adiabatic MSW effect for the neutrino flavor transformations in the supernova matter is also implemented. EstrellaNueva considers several supernova models simulated by the Core-Collapse Modeling Group at the Max Planck Institute for Astrophysics. In addition, the primary flux parameters can be considered constant in time, which is useful to study soft regions of the supernova neutrino spectrum. This software complements other tools available to the supernova physics community, with the surplus of its simplicity and flexibility. 

\section*{Acknowledgements}

The authors would like to thank Hans-Thomas Janka and the Core-Collapse Modeling Group at the Max Planck Institute for Astrophysics, for granting access to their database. The authors also thank Eduardo Peinado and Leon M. G. de la Vega from IF-UNAM and the Supernova group of the SNO+ collaboration, for useful discussions.\\

\noindent This work is supported by the projects CONACYT CB-2017-2018/A1-S-8960,  DGAPA UNAM grant PAPIIT-IN108020, and Fundación Marcos Moshinsky.

%% For this sample we use BibTeX plus aasjournals.bst to generate the
%% the bibliography. The sample631.bib file was populated from ADS. To
%% get the citations to show in the compiled file do the following:
%%
%% pdflatex sample631.tex
%% bibtext sample631
%% pdflatex sample631.tex
%% pdflatex sample631.tex

\bibliography{sample631}{}
\bibliographystyle{aasjournal}

%% This command is needed to show the entire author+affiliation list when
%% the collaboration and author truncation commands are used.  It has to
%% go at the end of the manuscript.
%\allauthors

%% Include this line if you are using the \added, \replaced, \deleted
%% commands to see a summary list of all changes at the end of the article.
%\listofchanges

\end{document}